\documentclass[12pt]{article}
\usepackage[utf8x]{inputenc} 
\usepackage{amssymb}
\usepackage{mathrsfs}
\usepackage{bm}
\usepackage{graphicx}
\usepackage{subfigure}
\usepackage{hyperref}
\usepackage{cite}
\begin{document}
\begin{titlepage}
\begin{center}
\begin{Large}
{\bf  A particle in equilibrium with a bath realizes worldline supersymmetry}
\end{Large}
\vskip1truecm
Stam Nicolis\footnote{E-Mail: stam.nicolis@lmpt.univ-tours.fr; stam.nicolis@idpoisson.fr}

{\sl Institut Denis Poisson, Université de Tours, Université d'Orléans, CNRS (UMR7013)\\
Parc Grandmont, Tours 37200, France}

\end{center}

\vskip1truecm

\begin{abstract}
We study the relation between the partition function of a non--relativistic particle, in one spatial dimension, that describes the equilibrium fluctuations implicitly, and the partition function of the same system, deduced from the Langevin equation, that describes the fluctuations explicitly, of a bath with additive white--noise properties using Monte Carlo simulations for computing the correlation functions that satisfy the corresponding identities.   We show that both can be related to the partition function of the corresponding, maximally supersymmetric, theory with one--dimensional bosonic worldvolume, by appropriate analytic continuation, from Euclidian to real time,  and that they can all describe the same physics, since the correlation functions of the observables satisfy the same identities for all systems.The supersymmetric theory provides the consistent closure for describing the fluctuations. 

Therefore supersymmetry is relevant at the scale in which equilibrium with the bath is meaningful. At scales when the ``true'' degrees of freedom of the bath can be resolved (e.g. atoms and molecules for the case of thermal fluctuations) the superpartners become ``hidden''. 
They can be, always, revealed through the identities satisfied by the correlation functions of 
the appropriate noise field, however. 

In fact, the same formalism  applies whatever the ``microscopic'' origin of the fluctuations. 

Therefore, all consistently closed physical systems are supersymmetric--and any system that is explicitly not invariant under supersymmetric transformations, is, in fact, open and, therefore, incomplete.
\end{abstract}
\end{titlepage}
\newpage

\section{Introduction}
The stochastic approach for describing the  dynamics of commuting degrees of freedom, in interaction with a bath,  starts with the Langevin equation in the presence of, additive, white noise
\begin{equation}
\label{langevin1}
\frac{\partial\phi(t)}{\partial t} = -\frac{\partial U(\phi)}{\partial\phi(t)}
+ \eta(t)
\end{equation}
where $t$ is the equilibration time. The field, $\eta(t)$, is a Gaussian
stochastic process:
\begin{equation}
\label{gaussian_noise}
\begin{array}{l}
\left\langle\eta(t)\right\rangle = 0\\
\left\langle\eta(t)\eta(t')\right\rangle = \nu\delta(t-t')\\
\left\langle
\eta(t_1)\eta(t_2)\cdots\eta(t_{2n})
\right\rangle
= \sum_{\pi}
\left\langle \eta(t_{\pi(1)})\eta(t_{\pi(2)})\right\rangle\cdots
\left\langle \eta(t_{\pi(2n-1)})\eta(t_{\pi(2n)})\right\rangle
\end{array}
\end{equation}
where the sum is over all permutations. The qualification ``additive'' means that the coefficient of the noise term does not depend on the dynamical variable(s), $\phi$ and can be taken as a constant, whose value can be set to 1.

At equilibrium, $t\to\infty$,
\begin{equation}
\eta = \frac{\partial U(\phi)}{\partial\phi}
\end{equation}
We haven't yet committed to whether $\eta$ and $\phi$ depend on other
variables; if they don't, we have a problem in probability theory of a finite number of degrees of freedom, that is
interesting, not only  in its own right, but can, also, provide insights for physics~\cite{nicolis_zerkak,Rossi:2010fc}.
 If they do, we have  a  problem in stochastic processes, the probability theory of an infinite number of degrees of freedom, where the noise process, $\eta$, models the  equilibrium fluctuations~\cite{parisi_sourlas,Feigelman:1983ac,Tailleur:2005su,ZinnJustin:2002ru}. Indeed, depending on how we interpret the coefficient, $\nu$, of the 2--point function of the stochastic field, $\eta$, this framework can describe quantum fluctuations, if the coefficient is proportional to $\hbar$, thermal fluctuations, if it is proportional to $k_\mathrm{B}T$, or disorder, if it is proportional to the ``strength of the noise'' (cf. ~\cite{Aron:2010ac} and~\cite{Nicolis:2016lof} for an example of multiplicative, colored, noise in this framework). In any given context, we can work in units where its value is set to 1. In the following we shall use language that's appropriate for the case where this scale is set by $\hbar$--but the mathematical treatment is completely general.

If $U(\phi)$ is a local functional of $\phi$, in particular, if 
\begin{equation}
\label{Uqm}
\frac{\partial U(\phi)}{\partial\phi} =\frac{\partial\phi(\tau)}{\partial\tau}
+ \frac{\partial W(\phi)}{\partial\phi(\tau)}
\end{equation}
where $\tau\in\mathbb{R}$ and $W(\phi(\tau))$ an ultralocal functional of
$\phi(\tau)$, we obtain the following stochastic equation for $\phi(\tau)$:
\begin{equation}
\label{langevin2}
\frac{\partial\phi(\tau)}{\partial\tau} = -\frac{\partial W}{\partial\phi(\tau)}+\eta(\tau)
\end{equation}
This is the Langevin equation that describes, for example, quantum mechanics, i.e. a quantum
field theory in one Euclidian dimension~\cite{zambrini}; or the thermodynamics of a classical string, since the Euclidian time is identified with the coordinate that labels the classical string (this has been stressed by Polyakov~\cite{Polyakov:1987ez}). 
 The essential difference to
eq.~(\ref{langevin1}) is that we are not interested, only, in the limit
$\tau\to\infty$, but in  the solution for all values of $\tau$. This assertion
is meaningful only at the level of the correlation functions, of course. 

In this case, $\eta(\tau)$ is a Gaussian stochastic process, whose correlation functions obey the same identities as in eq.~(\ref{gaussian_noise}), only the
time is, now, the Euclidian time. 

We are interested in the correlation functions,
$\left\langle\phi(\tau_1)\phi(\tau_2)\cdots\phi(\tau_n)\right\rangle$, of the
field $\phi$ and the identities that constrain them. This information can be obtained from the partition function of the noise, that describes the properties of the correlation functions of the noise and the Langevin equation, that defines the change of variables from the noise, $\left\{\eta(\tau)\right\}$ to the field $\{\phi(\tau)\}$~\cite{parisi_sourlas,ZinnJustin:2002ru,Nicolai:1980js,Nicolai:1980jc}. In equations:
\begin{equation}
\label{partition_function_phi}
\begin{array}{l}
\displaystyle
Z_\mathrm{L}= \int\,\left[{\mathscr D}\eta\right]\,e^{-\int\,d\tau\,\frac{1}{2}\eta(\tau)^2} = 1 = 
\int\,\left[{\mathscr D}\phi\right]\,\left|\mathrm{det}\left(\frac{\delta\eta(\tau)}{\delta\phi(\tau')}\right)\right|\,e^{-\int\,d\tau\,\frac{1}{2}\left(\dot{\phi}+W'(\phi)\right)^2}=\\
\displaystyle
\int\,\left[{\mathscr D}\phi\right]\,\mathrm{det}\left(\frac{\delta\eta(\tau)}{\delta\phi(\tau')}\right)\mathrm{sign}\left(\mathrm{det}\left(\frac{\delta\eta(\tau)}{\delta\phi(\tau')}\right) \right)
\,e^{-\int\,d\tau\,\frac{1}{2}\left(\dot{\phi}+W'(\phi)\right)^2}
\end{array}
\end{equation}
We assume, here, that the determinant is real, so that its phase is only 0 or $\pi$. If the phase can take other values, then the sign must be replaced by $\exp(\mathrm{i}\theta_\mathrm{det})$ here and in the following expressions. What is important is that, in $Z_\mathrm{L}$, only the absolute value of the determinant enters, so the integrand of  $Z_\mathrm{L}$ is, manifestly, real--and positive. 

The operator, whose determinant appears here is given by the, formal, expression
\begin{equation}
\label{detstoch}
\begin{array}{l}
\displaystyle
\frac{\delta\eta(\tau)}{\delta\phi(\tau')}=\delta(\tau-\tau')\left(\frac{d}{d\tau}+W''(\phi(\tau))\right)\Rightarrow \mathrm{det}\left(\frac{\delta\eta(\tau)}{\delta\phi(\tau')}\right)=\mathrm{det}\left(\delta(\tau-\tau')\left(\frac{d}{d\tau}+W''(\phi(\tau))\right)\right)=\\
\displaystyle
\hskip10.5truecm
\mathrm{det}\left(\delta(\tau-\tau')\right)\mathrm{det}\left(\frac{d}{d\tau}+W''(\phi(\tau))\right)
\end{array}
\end{equation}
We have introduced the subscript ``L'', for this partition function, in order to highlight that it is the partition function of the scalar field, that is constructed through the mapping of the Langevin equation. 

The determinant of the $\delta-$function contributes, upon regularization, a constant that's independent of any field configuration and, thus, may be absorbed in the normalization, without contributing to the dynamics. Its sign is, also, upon regularization, a global constant that can be absorbed in the normalization, also. The physics is to be found in the phase and the value of the determinant of the local operator, $d/d\tau+W''(\phi)$.

These are formal expressions, since the measure of integration is not defined and the determinant of the operator requires regularization. In addition, if it were possible for the expectation value, taken with respect to the, hitherto ill-defined measure, of the absolute value of the determinant to vanish, this would mean that the probability density, whose integral is the partition function, isn't normalizable and the correlation functions of the scalar would only be defined in the presence of a cutoff in the space of field configurations and depend on the details thereof. In the present case we do not expect this to happen, unless the classical potential is, either,  unbounded from below, or doesn't confine sufficiently rapidly at infinity, since, if it is an ultra--local functional of the scalar, that does possess both these properties, the interactions remain local and no phase transition can occur in a system with one-dimensional worldvolume--the potential obtained by taking into account all quantum corrections, always possesses a unique minimum.  This statement, however, requires further elaboration, in order to lead to an effective computational scheme for the correlation functions and their identities. We shall use a lattice regularization to define the partition function and Monte Carlo simulations to compute the correlation functions and the identities that describe the physics of this class of theories. 

A first insight now stems from the following fact: In the path integral formalism~\cite{Dirac:1933xn,Feynman:1948ur,Hartle:1991bb} to (Euclidian time) quantum mechanics, the partition function, of the same system, is given by the  expression
\begin{equation}
\label{ZQM}
Z_\mathrm{QM}\equiv\int\,\left[{\mathscr D}\phi\right]\,e^{-S[\phi]}
\end{equation}
where $S[\phi]$ is the classical action, given by the expressions
\begin{equation}
\label{Sclass}
\begin{array}{l}
\displaystyle
S[\phi]=\int\,d\tau\,\left(\frac{1}{2}\dot{\phi}^2+V(\phi)\right)=\mathrm{const}+\int\,d\tau\,\left(\frac{1}{2}\dot{\phi}^2+\frac{1}{2}W'(\phi)^2\right)=\\
\displaystyle
\hskip5.7truecm
\mathrm{const}\mp\int\,d\tau\,\dot{\phi}W'(\phi)+\int\,d\tau\,\frac{1}{2}\left(\pm\dot{\phi}+W'(\phi)\right)^2=\\
\displaystyle
\mathrm{const}+\int\,d\tau\,\frac{1}{4}\left[\left(\dot{\phi}+W'(\phi)\right)^2 + \left(-\dot{\phi}+W'(\phi)\right)^2\right] = \\
\displaystyle
\mathrm{const}+\int\,d\tau\,\frac{1}{4}\left(\pm\mathrm{i}\dot{\phi}+W'(\phi)\right)\left(\mp\mathrm{i}\dot{\phi}+W'(\phi)\right) =\\
\displaystyle
\mathrm{const}-\int\,d\tau\,\frac{1}{4}\left(\dot{\phi}\pm\mathrm{i}W'(\phi)\right)\left(-\dot{\phi}\pm\mathrm{i}W'(\phi)\right) 
\end{array}
\end{equation}
All these forms are of interest: The second and third, and further, lines highlight that, up to surface terms, the action is positive semi-definite and displays the ``Bogomolny'i'' bound--that's saturated by solutions of first order equations~\cite{Mezincescu:2014zba,Mezincescu:2015apa,Baulieu:1993zm}.

These expressions are valid for {\em any} potential, $V(\phi)$, that's bounded from below and confines, sufficiently rapidly,  at infinity; by adding the appropriate constant to it, it can be written as a non--negative quantity, namely the square of a function, $W'(\phi)$, of the scalar, that's manifestly real. Of course $W'(\phi)$ need not be a polynomial, if $V(\phi)$ possesses multiple minima--but this is not relevant for the validity of these expressions. On the other hand, if $W'(\phi)$ is a polynomial, so is $V(\phi)$. 

If the classical potential isn't bounded from below, or doesn't confine at infinity, then the partition function isn't well--defined, for the scattering states, that describe the escape from the classical potential well through tunneling, or possess energy greater than the height of the potential, are not taken into account by it and it must be ``completed'', by constructing the potential for the scattering states. How to do this in the stochastic formalism  is beyond the scope of the present paper.

What's relevant is that the two partition functions, $Z_\mathrm{L}$ and $Z_\mathrm{QM}$, are supposed to describe the same system, if the assumption that the noise term of the Langevin equation describes the fluctuations is correct--and that they provide a complete description. 
This means that the correlation functions computed with either must be equal. We notice that there is one difference between them: $Z_\mathrm{L}$ contains the absolute value of the determinant of a local operator, that described the change of variables between noise and scalar, $Z_\mathrm{QM}$ doesn't contain it, explicitly.  However, the two can be related by noting that 
\begin{equation}
\label{QM_Langevin}
Z_\mathrm{L}=Z_\mathrm{QM}\left\langle\left|\mathrm{det}\frac{\delta\eta}{\delta\phi}\right|\right\rangle_\mathrm{QM}=
Z_\mathrm{QM}\left\langle\left|\mathrm{det}\left(\frac{d}{d\tau}+W''(\phi(\tau))\right)\right|\right\rangle_\mathrm{QM}
\end{equation}
which implies that  the identity between correlation functions, of any observable ${\mathcal O}[\{\phi\}]$  can be expressed by an identity for the correlation functions computed from $Z_\mathrm{QM}$:
 \begin{equation}
 \label{First_identity}
 \begin{array}{l}
\displaystyle
 \left\langle{\mathcal O}[\{\phi\}]\right\rangle_\mathrm{L}=\left\langle{\mathcal O}[\{\phi\}]\right\rangle_\mathrm{QM} \end{array}
 \end{equation}
The equality of the partition functions, therefore, relies on the property that 
\begin{equation}
\label{First_identity_1}
\left\langle\left|\mathrm{det}\left(\frac{d}{d\tau}+W''(\phi(\tau))\right)\right|\right\rangle_\mathrm{QM} =1
\end{equation} 
More precisely, it relies on the property that the LHS doesn't depend on the field configuration and, therefore, may be taken  equal to 1, by an appropriate choice of units.

Furthermore, by introducing the partition function $Z_\mathrm{SUSY}$, defined by 
\begin{equation}
\label{Zsusy}
Z_\mathrm{SUSY}\equiv\int\,[\mathscr{D}\phi][\mathscr{D}\psi][\mathscr{D}\chi]\,e^{-S[\phi,\psi,\chi]}
\end{equation}
we may express 
 $Z_\mathrm{QM}$ and $Z_\mathrm{L}$ through
\begin{equation}
\label{ZLQMSUSY}
\begin{array}{l}
\displaystyle
Z_\mathrm{L}=\int\,[\mathscr{D}\phi]\,\left|\mathrm{det}\,\left(\frac{d}{d\tau}+W''(\phi)\right)\right|\,e^{-S[\phi]}=\\
\displaystyle\hskip1truecm
\int\,[\mathscr{D}\phi]\,\mathrm{sign}\left\{ 
\mathrm{det}\,\left(\frac{d}{d\tau}+W''(\phi)\right)
\right\}\mathrm{det}\,\left(\frac{d}{d\tau}+W''(\phi)\right)\,e^{-S[\phi]}=\\
\displaystyle\hskip1truecm
\left\langle \mathrm{sign}\left\{ 
\mathrm{det}\,\left(\frac{d}{d\tau}+W''(\phi)\right)
\right\}\right\rangle_\mathrm{SUSY}Z_\mathrm{SUSY}=\\
\displaystyle\hskip1truecm
\left\langle \mathrm{sign}\left\{ 
\mathrm{det}\,\left(\frac{d}{d\tau}+W''(\phi)\right)
\right\}\right\rangle_\mathrm{SUSY}\left\langle\mathrm{det}\,\left(\frac{d}{d\tau}+W''(\phi)\right)\right\rangle_\mathrm{QM}Z_\mathrm{QM}
\end{array}
\end{equation}
The first factor in the last two lines is the ``Witten index''~\cite{Witten:1981nf,Witten:1982im,Witten:1982df}.
It has been the object of considerable study--along with its ``refinements''. Since its topological properties were--and are--of interest in mathematics, it is the subject of continuing attention; however they aren't sufficient to constrain the dynamics, as fully as it was, initially, hoped.

The equivalence can be, in principle,  tested by performing calculations and experiments using the partition function of the path integral formalism of ordinary quantum mechanics~\cite{Dirac:1933xn,Feynman:1948ur,Hartle:1991bb}and involves an, apparently, unexpected identity. Since the determinant of a local operator is a non--local quantity, realizing this test is quite non--trivial, so it's useful to try and understand what the equality of the two partition functions really means, in order to look for identities, that are easier to test. This is what we shall present in what follows.  The work of refs.~\cite{parisi_sourlas,Nicolai:1980js,Nicolai:1980jc} showed that it is possible  to test the validity of the equivalence by studying the correlation functions of the noise, $\eta(\tau)$, which is defined, by the Langevin equation, in terms of the scalar, $\phi(\tau)$:
\begin{equation}
\label{Nicolai_map}
\eta(\tau)\equiv\eta[\phi(\tau)]=\frac{d\phi}{d\tau}+W'(\phi)
\end{equation}
This is called the ``Nicolai map'' in the literature~\cite{parisi_sourlas,Nicolai:1980js,Nicolai:1980jc,Damgaard:1983ys,Damgaard:1983tq} and can be considered as the definition of the field $\phi(\tau)$, in terms of the Gaussian field $\eta(\tau)$.  In principle, the solutions of this differential equation, are the field configurations of $\{\phi\}$ and there has been considerable effort in trying to make this correspondence practical~\cite{Beccaria:1998vi,Morikawa:2018ops,Morikawa:2018zys,Morikawa:2019cql}. 

This statement, however, is true, only in the case where $W''(\phi(\tau))$ is of fixed sign, i.e.  $V(\phi)$, has a unique minimum. If $W''(\phi)$ can vanish, then the classical potential has multiple minina. In the case at hand this is an expression of the classical limit and the semi--classical approximation: Quantum fluctuations lead to tunneling between the classical minima and the quantum potential has a unique minimum.  Therefore any numerical calculation of this identity, inevitably, checks the accuracy of the algorithm, to computing the exponentially small effects of tunneling. First attempts to treat them were reported in~\cite{Cecotti:1981fu,Cecotti:1982gn,Cecotti:1982ad,Cecotti:1983up}. These papers, also, adapted the opposite approach to that used here, in that they assumed supersymmetry, as an additional property. 

Therefore, if tunneling is suppressed, the equivalence between $Z_\mathrm{L}$ and $Z_\mathrm{QM}$ breaks down and the corrections, computed with $Z_\mathrm{QM}$, don't get repackaged into the absolute value of the determinant of the original field(s) and new degrees of freedom are required to achieve this. What these might be remains to be elucidated. 
Typical cases where tunneling is suppressed is across phase transitions; so we can expect that theories, above their lower critical dimension and below their upper critical 
dimension, display fluctuations that aren't captured by $Z_\mathrm{QM}$ (cf. \cite{Fytas:2019zdk} for a recent investigation in this context).

While the degrees of freedom themselves may be hard to identify, their presence can be revealed by the ``anomalies'' in the identities that the correlation functions satisfy.

Furthermore, while it's not possible to invert the Nicolai map, in any useful way, beyond perturbation theory, in general, it's, on the contrary, straightforward, as we shall show in the following, to compute the correlation functions of $\eta[\phi(\tau)]$ and the identities that constrain them by using the partition function $Z_\mathrm{QM}$--by Monte Carlo simulations, and not by Simpson's rule (or any other direct approach), since we are dealing with $N-$dimensional integrals, using a lattice regularization of the classical action, $S[\phi]$, that's perfectly well behaved. 

This calculation isn't trivial for the following reasons: if we use $Z_\mathrm{L}$, we must compute the correlation functions of $\eta[\phi(\tau)]$, in the presence of the insertion of the non--local determinant, that, in addition, has zeromodes to be accounted for, when $W''(\phi)$ can vanish. While the presence of the determinant ensures that their correlation  functions do satisfy Wick's theorem, what happens when $W''(\phi)$ can vanish isn't obvious. 
 If we use $Z_\mathrm{QM}$, things are more interesting: then it isn't at all obvious that the correlation functions of $\eta[\phi(\tau)]$, sampled without the explicit presence of the phase of the determinant, do satisfy Wick's theorem--it's the result of the fact that by sampling the configuration space in a way that takes into account the quantum fluctuations, the contribution of the determinant is, implicitly, taken into account. This by no means obvious and is a non--trivial statement about the fluctuations, that can be of relevance in many experimental situations, with ion traps or ultrafast magnetization dynamics.  In this way we can treat all cases of potentials bounded from below, and confining at infinity, the only difference being whether  the 1--point function of the noise vanishes, or not; whether the 2--point function is ultra--local and whether the multi--point functions do satisfy Wick's theorem.

 This brings us to the issue of boundary conditions. We shall study, in this paper, the case of periodic boundary conditions, the study of other boundary conditions will be reported in future work.   (It is interesting to note that, for the one--dimensional worldvolume studied here, periodic boundary conditions on the anticommuting fields imply that, when $Z_\mathrm{SUSY}$ is well--defined, since $Z_\mathrm{SUSY}\equiv 1$, then $Z_\mathrm{QM}=$Witten index~\cite{AlvarezGaume:1983at}.) 
 
 Periodic boundary conditions for the scalar  would imply that $\langle\dot{\phi}(\tau)\rangle = 0$. This has the following consequence: Since $\left\langle\eta[\phi(\tau)]\right\rangle=0=\langle\dot{\phi}\rangle+\left\langle W'(\phi)\right\rangle$, we realize that, if $\left\langle W'(\phi)\right\rangle= 0$, then $\langle\dot{\phi}\rangle= 0$, also and the translation invariance, that's allowed by the boundary conditions, is, indeed, realized by the dynamics. If $\left\langle W'(\phi)\right\rangle\neq 0$, then it might seem that $\langle\dot{\phi}\rangle\neq 0$ and the translation invariance, that's allowed by the boundary conditions, is not realized by the dynamics, if $\left\langle\eta[\phi(\tau)]\right\rangle=0$.  In this case the system  isn't closed, but remains in equilibrium, in similar fashion to a system subject to an ``external'', stationary, field is in equilibrium, but not closed.  
 
  This is the new feature, in this case and expresses the fact that the ``quantum'' potential has completely different properties from the ``classical'' potential, the ``backreaction'' of the fluctuations on the potential cannot be neglected and changes its form completely--and can be an example of a situation where it isn't, in fact, possible to integrate out the fermions, at fixed configuration of the scalar~\cite{Witten:2015aba,Bergshoeff:2014gja}, while neglecting the backreaction--it is, possible, however, when taking it into account, as is done by the computation of the correlation functions, by Monte Carlo simulations. It is this result that expresses, in another, though, of course, equivalent, way, the breakdown of the dimensional reduction, $D\to D-2$~\cite{parisi_sourlas}. The same reasons hold for the investigations of refs.~\cite{Cardy:1983aa,Kirschner:1984ms,Kirschner:1984sx}, that focused, also,  on this aspect; as well as those that relied  on the assumption that supersymmetry is discretionary~\cite{Salomonson:1981ug,Macfarlane:1997wa,vanHolten:1995bw,deAzcarraga:2001jdk,Horvathy:2010vm}. 
  
It should be stressed, however, that the previous statements refer to uniform configurations and don't cover the case of non--uniform configurations, that would satisfy 
$\langle\dot{\phi}+W'(\phi)\rangle=0$, without satisfying $\langle\dot{\phi}\rangle=0=\langle W'(\phi)\rangle$, if supersymmetry is unbroken. Indeed, we remark that non--uniform configurations, that satisfy $\langle\dot{\phi}+W'(\phi)\rangle=0$  can describe supersymmetric configurations, even though $\langle W'(\phi)\rangle\neq 0$. This would mean, of course, that (Euclidian) translation invariance is spontaneously broken, $\langle\dot{\phi}\rangle\neq 0$. More precisely, 
(Euclidian) time translation invariance isn't, in this case, a global symmetry, but, only, a local one: The target space isn't, globally, flat. 
 
In the present paper we focus on uniform configurations and defer the case of non--uniform configurations to future work.   
     
We have expressly presented the subject in a way that stresses that these issues are relevant in general and can be treated this way. Now, of course, it's appropriate to confirm that appearances are not deceiving: there is a ``deeper''  explanation. If we introduce into the classical action the determinant of the operator $d/d\tau+W''(\phi)$, as a local contribution, we must use anticommuting fields. Then  we can show that the full action is  invariant under transformations, with anticommuting parameters,  that map commuting fields to anticommuting fields and vice versa and close on the translations, therefore deserve to be called ``supersymmetric''~\cite{parisi_sourlas,Nicolai:1980js,Nicolai:1980jc,deAlfaro:1982ex,deAlfaro:1984gw,deAlfaro:1984hb,DeAlfaro:1986uv}. However it is by no means obvious that the fluctuations, computed through  the partition function, $Z_\mathrm{L}$, respect this symmetry. A major technical issue is that the phase of the determinant has remained and this, too, of course, is a non--local quantity, whose treatment isn't at all straightforward. The difficulty in dealing with the presence of the phase of the determinant is known as ``the sign problem'' and there has been considerable work done in trying to solve it, in order to sample the field configurations with $Z_\mathrm{L}$ efficiently~cf., for instance~\cite{Kastner:2007gz,Wozar:2011gu,Schierenberg:2012pb,Bergner:2009vg,Bergner:2012nu,Baumgartner:2015zna,Baumgartner:2015qba,Baumgartner:2014nka,Baumgartner:2013ara,Baumgartner:2011jw,Baumgartner:2011cm,Baumgartner:2012np,Bergner:2016qbz,Bergner:2015ywa,Catterall:2010nd,Catterall:2009it,Catterall:2015ira,DAdda:2015fim,DAdda:2010hbr,DAdda:2009fqg,Arianos:2007nv,DAdda:2004ypx}.

However, we realize that we can check the equivalence of the partition functions, by computing the correlation functions using $Z_\mathrm{QM}$, that only involves local quantities and commuting fields. The anticommuting fields are ``hidden'', as is the determinant (along with its phase) in their  fluctuations. Their presence can be revealed by appropriate probes, namely, the correlation functions of $\eta[\phi(\tau)]$. If we can show that the correlation functions of the field, $\eta[\phi(\tau)]$, as functionals of the scalar, do satisfy Wick's theorem, when the scalar field is sampled using $Z_\mathrm{QM}$, we will have provided evidence,  that the fluctuations are, indeed, described by the contribution of the determinant;  also, when $W''(\phi)$ may vanish, in fact and supersymmetry is broken. 

The physical meaning of these statements is that the Langevin equation and the partition functions, $Z_\mathrm{L}$ and $Z_\mathrm{QM}$, describe the dynamics of a non--relativistic particle, whose trajectory is defined by the scalar field, $\phi(\tau)$, in contact with a bath, that describes the fluctuating ``environment'', quantum, thermal or disorder. The Langevin equation leads to the, explicit, description of the bath in terms of local, but {\em anticommuting} degrees of freedom--in contrast to stochastic formalisms, that treat the bath in terms of commuting degrees of freedom, that, inevitably,  have non--local dynamics.  The partition function of the path integral formalism, $Z_\mathrm{QM}$, takes into account the fluctuations implicitly: either perturbatively, through the loop expansion, or non--perturbatively, through the sampling of the configuration space of the fields by Monte Carlo methods. In both cases it is possible, in principle, to compute the identities that express the fact that the system particle+bath is closed and that it is probed consistently by specifying an action that has a physical kinetic term and a  potential that's bounded from below and confines at infinity. And the geometrical description is provided by superspace (cf. also~\cite{Gozzi:1991wi} and~\cite{Kugo:1989xs}). 

On the other hand, while the anticommuting variables (assuming the dynamics is described by commuting variables) can resolve the fluctuations and can evade the assumptions of Bell's theorem~\cite{Freund:1981yp,Dzhunushaliev:2007vg}, they don't play the role of hidden variables, as envisaged by Einstein, since the partition function 
doesn't become a $\delta-$function(al) on the solutions of the classical equations of motion of the supersymmetric theory. 

The plan of the paper is the following:

In section~\ref{SUSY_class} we recall that the classical action in $Z_\mathrm{L}$, when taking into account the determinant,  is invariant under transformations that map the commuting field, $\phi(\tau)$, to an anticommuting field and vice versa. The infinitesimal parameters are anticommuting variables and the algebra defined by these transformations is shown to close on the translations in Euclidian time, thereby justifying calling these transformations supersymmetric. This confirms that the symmetry presented in refs.~\cite{parisi_sourlas,Nicolai:1980js,Nicolai:1980jc} is, indeed, a supersymmetry and provides fresh motivation for experimental searches, that were not possible thirty years ago. We, also, remark the existence of a, global, $SO(1,1)$, symmetry that acts, only, on the fermions~\cite{Zumino:1977yh}. 

We  deduce the identities that the correlation functions must satisfy, if the partitions functions, $Z_\mathrm{QM}$ and $Z_\mathrm{L}$, in fact, are equal to the partition function of a manifestly supersymmetric theory, described by the partition function $Z_\mathrm{SUSY}$. These identities have consequences that can be checked by computing correlation functions of the noise field, $\eta[\phi(\tau)]$, with respect to $Z_\mathrm{QM}$. It will be these consequences that we will explore by the lattice formulation of $Z_\mathrm{QM}$. From the point of view of geometry, the bottom line is that worldvolume and target space become non--commutative manifolds, when fluctuations are taken into account and their structure reflects the fact that fluctuations are consistently described--that backreaction is fully taken into account. In this way it becomes clear that any non--local properties of the observables may be understood by the fact that locality makes sense, in general,  in the full manifold, not along any submanifold, which the commuting part becomes, when fluctuations must be taken into account. The only issue is whether the anticommuting fields, that provide the consistent description, are, themselves, fundamental, in this sense--and the non--trivial statement is that, for thermal and disorder fluctuations the answer is that they are, indeed, emergent; for quantum fluctuations, on the other hand, the answer is that they are, indeed, fundamental and cannot be reproduced by other, local, degrees of freedom, within the context of non--relativistic dynamics--this is the substance of the experimental tests of Bell's inequalities. Whether it's useful to describe them with non-local degrees of freedom is, of course, an open issue (cf. for instance~\cite{Bars:2014jca}).

In section~\ref{SUSY_qu} we set up the lattice formulation for computing correlation functions with $Z_\mathrm{QM}$ and discuss how the consequences can be probed by the lattice regularization, in particular, the  properties of the backreaction. 
 
For the case of the quartic and sextic anharmonic oscillators that are the focus of our study,
we show that, while the scalar potential of the quartic oscillator can, always, be written as the 
square of a polynomial in the field, up to an additive constant, this isn't the case for 
the sextic oscillator. This means that, in general, the superpotential for the sextic oscillator
isn't a polynomial in the field. 

In section~\ref{numerics} we present and discuss our numerical results, for the two cases. We use the quartic and cubic polynomial superpotentials as examples. We focus on the 1-- and 2--point functions and on the Binder cumulant, the ultra--local part of the connected 4--point function for the  noise field, expressed in terms of the scalar.

Our conclusions and ideas for further investigations are the topic of section~\ref{conclusions}.

For ease of exposition, we shall refer to the commuting fields, also,  as ``bosons'' and the anticommuting fields, also, as ``fermions''--they are target space scalars here, since the (bosonic part of the) target space is one--dimensional and the system is non--relativistic. A summary may be find in ref.~\cite{Nicolis:2016osp}. 

The extension of the present approach  to infinitely many particles, interacting in a way consistent with worldvolume {\em and} target space supersymmetry, namely the 1+1--dimensional Wess-Zumino model,is presented in ref.~\cite{Nicolis:2017lqk}.

\section{On the worldline supersymmetry of the classical action}\label{SUSY_class}
\subsection{The subtleties and symmetries of the fermionic action}\label{fermionic_action}
Let us start with the partition function, $Z_\mathrm{L}$, deduced from the Langevin equation,
\begin{equation}
\label{ZLrevisited}
Z_\mathrm{L}=\int\,\left[{\mathscr D}\phi\right]\,\mathrm{det}\left(\frac{d}{d\tau}+W''(\phi)\right)\mathrm{sign}\left(\mathrm{det}\left(\frac{d}{d\tau}+W''(\phi)\right) \right)
\,e^{-\int\,d\tau\,\frac{1}{2}\left(\dot{\phi}+W'(\phi)\right)^2}
\end{equation}
We want to introduce the determinant of the local operator in the action, that is local in the commuting field $\phi(\tau)$, in a way that respects this property. The way to achieve this is by introducing {\em two} anticommuting fields, $\psi(\tau)$ and $\chi(\tau)$. These are one--component quantities at each point of the worldline. Just like $\phi(\tau)$ has the property that $[\phi(\tau),\phi(\tau')]=0$, for all $\tau$ and $\tau'$, these have the properties that $\{\psi(\tau),\psi(\tau')\}=0=\{\chi(\tau),\chi(\tau')\}$ and $\{\psi(\tau),\chi(\tau')\}=0$. For the moment the singularities that can arise, as $|\tau-\tau'|\to 0$ are not taken into account--we shall encounter some of their effects further on. 

Then it can be shown that 
\begin{equation}
\label{detanticomm}
\mathrm{det}\left(\frac{d}{d\tau}+W''(\phi)\right)=\int\,[{\mathscr D}\psi(\tau)][{\mathscr D}\chi(\tau)]\,e^{\int\,d\tau\,\psi(\tau)\left(\frac{d}{d\tau}+W''(\phi)\right)\chi(\tau)}
\end{equation}
This equality can be motivated by working through finite dimensional examples: it's not possible to express the determinant using one anticommuting field only: If ${\sf M}$ is an $r-$dimensional, non--singular, matrix, then 
\begin{equation}
\label{finitedet}
\mathrm{det}\,{\sf M}=\int\,\prod_{I=1}^r\,d\psi_I\,d\chi_I\,e^{\psi_I{\sf M}^{IJ}\chi_J}\neq 
\int\,\prod_{I=1}^r\,d\psi_I\,e^{\frac{1}{2}\psi_I{\sf M}^{IJ}\psi_J}
\end{equation}
So we realize that this doubling of anticommuting degrees of freedom doesn't have anything to do with lattice effects, but is a property of the anticommuting variables themselves, a mathematical statement, that is the consequence of attempting to define the determinant of the matrix ${\sf M}$.

However, there's more to be said, for the properties of the exponent, for it can be written in a way that renders manifest that the two sets of anticommuting variables can mix, assuming the physics of the problem allows it. We may group the two sets in terms of two--component fields,
 $\Psi_A=\Psi_{\alpha,I}\equiv\left(\psi,\chi\right)_I^\mathrm{T}$, with $\Psi_{1,I}=\psi_I$ and $\Psi_{2,I}=\chi_I$; then we have the identity
\begin{equation}
\label{doubling}
\psi_I{\sf M}^{IJ}\chi_J=\frac{1}{2}\Psi_{\alpha,I}\varepsilon^{\alpha\beta}{\sf M}^{IJ}\Psi_{\beta,J}\equiv\frac{1}{2}\Psi_A{\sf N}^{AB}\Psi_B
\end{equation}
where the convention is 
\begin{equation}
\label{ }
\varepsilon^{\alpha\beta}=\left(\begin{array}{cc} 0 & -1\\1 & 0\end{array}\right)
\end{equation}
Therefore eq.~(\ref{finitedet}) becomes 
\begin{equation}
\label{finitedef}
\mathrm{det}\,{\sf M}=\sqrt{\mathrm{det}\,{\sf N}}
\end{equation}
The sign ambiguity of the square root is a global sign, that can be fixed, once and for all and doesn't have any physical significance; it has to do with the convention of where's the sign in the Levi--Civita tensor and in what order are the multiple integrations over the anticommuting variables carried out--it is totally distinct from the question of the sign--more generally, the phase--of the determinant of {\sf M}, that is of physical significance. Recall that we have explicitly factored out the phase  of the determinant of the local operator, when we introduced the fermions, so we're, for the moment, implicitly, assuming the sign is constant. We discuss below how to test this assumption.

For the differential operator that's of interest here, one can draw the following conclusions: First of all,  that one of the anticommuting  fields isn't dynamical, but enforces a constraint, since it enters in the Lagrangian without a time derivative and, by partial integration, it's possible to choose, which field of the two appears with its time derivative and which doesn't. Next, that one may write the Lagrangian in the following way, that emphasizes a more ``symmetric'' treatment of the two fields, in terms of the combinations $\Psi_A(\tau)$. We find that the potential term can be written as 
\begin{equation}
\label{mass_term}
W''(\phi)\psi(\tau)\chi(\tau)=\frac{1}{2}W''(\phi)\left(\psi(\tau)\chi(\tau)-\chi(\tau)\psi(\tau)\right)=\frac{1}{2}W''(\phi)\left[\psi(\tau),\chi(\tau)\right]=
\frac{1}{2}W''(\phi)\Psi_A\varepsilon^{AB}\Psi_B
\end{equation}
The kinetic term is more interesting:
\begin{equation}
\label{kinetic_term}
\psi(\tau)\frac{d}{d\tau}\chi(\tau)=\frac{1}{2}\left(\psi\dot{\chi}+\dot{\psi}\chi\right)+\frac{1}{2}\left(\psi\dot{\chi}-\dot{\psi}\chi\right)
\end{equation}
The first term is a total derivative, while the second term is not.  In terms of the $\Psi_A(\tau)$ it becomes 
\begin{equation}
\label{kinetic_term_A}
\frac{1}{2}\left(\psi\dot{\chi}+\dot{\psi}\chi\right)+\frac{1}{2}\left(\psi\dot{\chi}-\dot{\psi}\chi\right) =\frac{1}{2}\Psi_A\sigma_1^{AB}\dot{\Psi}_B-\frac{\mathrm{i}}{2}\Psi_A\sigma_2^{AB}\dot{\Psi}_B=\Psi_A\sigma_-^{AB}\dot{\Psi}_B
\end{equation}
Here there's a sign ambiguity: depending on the sign of the surface term, it's possible to obtain either $\sigma_+$ or $\sigma_-$. What's interesting to note is that the $\sigma_\pm$ are ladder operators. It's the ladder property that ensures that the correct number of degrees of freedom propagates, despite the redundancy introduced by the inevitable doubling of the degrees of freedom. Therefore one must specify the appropriate vacuum, on which they act as creation operators: any one of the two will do, but one must be specified. This is what the boundary conditions, including the boundary terms, accomplish. In addition, we note the presence of  a global, $SO(1,1)$ symmetry~\cite{Zumino:1977yh}, 
\begin{equation}
\label{SO11}
\begin{array}{ccc}
\displaystyle
\delta_\theta\psi=\theta\psi & \mathrm{and} & \displaystyle \delta_\theta\chi= -\theta\chi
\end{array}
\end{equation}
that acts only on the fermions, $\delta_\theta\phi=0$.  What its precise consequences, in the present context, are, remains to be clarified. It appears that the two fermions don't mix under this transformation.

The potential term is, of course, invariant, also, under it if it becomes promoted to a local symmetry, the kinetic term, however, is not. 

The bottom line  is that anticommuting degrees of freedom, inevitably, lead to a redundant description of the dynamics and this occurs, in the continuum and, even, when they cannot be identified with target space fermions, since, in a non--relativistic context, there isn't any spin--statistics theorem~\cite{Wightman2000}. This redundancy  isn't an artifact of any regularization scheme, but reflects a mathematical property of anticommuting variables and fields. While its consequences have been described in string theory~\cite{Gliozzi:1976qd,Gliozzi:1976jf,Grassi:2011ie}, its relevance for field theory has received attention only as expressing lattice artifacts; it certainly deserves a closer look (cf.~\cite{Catterall:2010nd,Catterall:2009it,Catterall:2015ira,DAdda:2015fim,DAdda:2010hbr,DAdda:2009fqg,Arianos:2007nv,DAdda:2004ypx})

In these considerations, the fact that the scalar field could propagate didn't matter. 

In the next section we shall  study what happens when the scalar field, that appears in the fermionic Lagrangian through the ``Yukawa'' term $\psi W''(\phi)\chi$ can propagate, i.e. it can become dynamical. Then a new global symmetry appears, that mixes the commuting and the anticommuting fields.  

\subsection{Worldline supersymmetry and its algebra}\label{wlSUSY}
When the scalar field can become dynamical, i.e. appears with its kinetic term, the full action, defined by $Z_\mathrm{L}$,  is given by the following expression,
\begin{equation}
\label{fullQM}
\begin{array}{l}
\displaystyle
S[\phi,\psi,\chi]=\int\,d\tau\,\left\{ 
\frac{1}{2}\dot{\phi}^2+\frac{1}{2}W'(\phi)^2-\frac{1}{2}\left(\psi\dot{\chi}-\dot{\psi}\chi\right)-W''(\phi)\frac{1}{2}\left(\psi\chi-\chi\psi\right)
 \right\} = \\
 \displaystyle
 \hskip2truecm
 \int\,d\tau\,\left\{
 \frac{1}{2}\dot{\phi}^2-\frac{F^2}{2}+FW'(\phi)-\frac{1}{2}\left(\psi\dot{\chi}-\dot{\psi}\chi\right)-\frac{1}{2}W''(\phi)\left[\psi,\chi\right]
 \right\} 
 \end{array}
\end{equation}
where we have dropped the surface term that arises from the kinetic term of the fermions. 

We have used an auxiliary field, $F(\tau)$ (that's integrated along the imaginary axis), to linearize the potential term for the scalar. We remark that there is a perfect matching between the number of commuting and anticommuting fields, namely, two of each. As we discussed above, one anticommuting degree of freedom  is an artifact--and, here, we see that one, commuting, degree of freedom is, also,  an artifact.  So the matching of ``physical'' degrees of freedom is, also, realized. 

The question, then, arises, whether this matching is just an accident, or whether there is a symmetry that describes and controls it. If there is such a symmetry, it is by no means inevitable that it is a supersymmetry--there do exist such degeneracies between the spectra of fermions and bosons that are described by symmetries that do not close on the translations (an example is the ``massive fermion--boson degeneracy''~\cite{Florakis:2010ty}, that may be relevant, also, in the present context, where only massive particles have well--defined dynamics, since the dynamics is non--relativistic). 

The work of refs.~\cite{deAlfaro:1982ex,deAlfaro:1984gw,deAlfaro:1984hb,DeAlfaro:1986uv,Cooper:1982dm,Bender:1983xi,Freedman:1983as,deLimaRodrigues:2002ya} indicates that this action is, indeed, invariant under supersymmetric transformations. For, if we consider  the transformations, defined by operators, $Q_1$ and $Q_2$,
 \begin{equation}
\label{SUSYalgebra}
\begin{array}{lcl}
\displaystyle
Q_1\phi = -\chi & & \displaystyle Q_2\phi = \psi\\
\displaystyle
Q_1\chi = 0 & &\displaystyle Q_2\chi = \dot{\phi}-F\\
\displaystyle 
Q_1\psi = -\dot{\phi}-F & &\displaystyle Q_2\psi = 0\\
\displaystyle 
Q_1 F = \dot{\chi} & & \displaystyle Q_2 F = \dot{\psi}
\end{array}
\end{equation}
we can easily deduce that $Q_1^2=0=Q_2^2$ and that their anticommutator closes on the translation operator in Euclidian time,
\begin{equation}
\label{anticommQ1Q2}
\left\{Q_1,Q_2\right\}=-2\frac{d}{d\tau}\Leftrightarrow\left[\zeta_1Q_1,\zeta_2Q_2\right]=2\zeta_1\zeta_2\frac{d}{d\tau}=-\zeta_\alpha\varepsilon^{\alpha\beta}\zeta_\beta\frac{d}{d\tau}
\end{equation}
The relevance of these transformations is that they're symmetries of the classical action, i.e. that $\delta_1 S\equiv\zeta_1 Q_1 S =0$ and $\delta_2 S\equiv \zeta_2Q_2 S=0$, up to total derivatives:
\begin{equation}
\label{Sinvariant}
\begin{array}{l}
\displaystyle
\delta_1 S_\mathrm{free}\equiv \int\,d\tau\,\left\{\dot{\phi}\zeta_1Q_1\dot{\phi}-F\zeta_1Q_1F-\left(\zeta_1Q_1\psi\right)\dot{\chi} \right\}=\zeta_1\int\,d\tau\,\left\{ -\dot{\phi}\dot{\chi}-F\dot{\chi}+\dot{\phi}\dot{\chi}+F\dot{\chi}\right\}=0\\
\displaystyle
\delta_2 S_\mathrm{free}\equiv \int\,d\tau\,\left\{\dot{\phi}\zeta_2Q_2\dot{\phi}-F\zeta_2Q_2F-\psi\left(\zeta_2Q_2\dot{\chi}\right) \right\}=\zeta_2\int\,d\tau\,\left\{\dot{\phi}\dot{\psi}-F\dot{\psi} +\psi\left( \ddot{\phi}-\dot{F}\right)\right\}=\\
\displaystyle
\hskip1truecm\zeta_2\int\,d\tau\,\frac{d}{d\tau}\left\{\left(\dot{\phi}-F\right)\psi\right\}\\
\displaystyle
\delta_1 S_\mathrm{int}\equiv \int\,d\tau\,\left\{\zeta_1\left(Q_1F \right)W' + \zeta_1 F W''Q_1\phi-\zeta_1\left( Q_1\psi\right)W''\chi \right\}=\\
\displaystyle
\hskip1truecm
\zeta_1\int\,d\tau\,\left\{ W'\dot{\chi}-FW''\chi+\left(\dot{\phi}+F\right)W''\chi\right\}=\zeta_1\int\,d\tau\,\frac{d}{d\tau}\left(W'\chi\right)\\
\displaystyle
\delta_2 S_\mathrm{int}\equiv\int\,d\tau\,\left\{ \zeta_2\left(Q_2F \right)W' + \zeta_2 F W''Q_2\phi+\zeta_2\psi W''\left(Q_2\chi\right)  \right\} = \\
\displaystyle
\hskip1truecm
\zeta_2\int\,d\tau\,\left\{ \dot{\psi}W' + FW''\psi +\psi W''\dot{\phi}-\psi W'' F\right\}=\zeta_2\int\,d\tau\,\frac{d}{d\tau}\left(W'\psi\right)
\end{array}
\end{equation}
What's important is that the parameters, $\zeta_{1,2}$, of the transformations anticommute with the fields $\psi(\tau)$ and $\chi(\tau)$ and commute with the fields $\phi(\tau)$ and $F(\tau)$.

In the above equations, the reason we found $\delta_1 S_\mathrm{free}=0$ identically is because the $\psi$ field didn't enter with a kinetic term, it imposed a constraint.  Adding the contribution of the boundary term, as in  eq.~(\ref{fullQM})
\begin{equation}
\label{delta1Sfree_boundary}
\begin{array}{l}
\displaystyle
\delta_1 S_\mathrm{free}^{(b)}=\int\,d\tau\,\frac{1}{2}\left(\left(\delta_1\psi\right)\dot{\chi}+\left(\delta_1\dot{\psi}\right)\chi\right) =\zeta_1\frac{1}{2}\int\,d\tau\,\left\{ \left(Q_1\psi\right)\dot{\chi}+\left(Q_1\dot{\psi}\right)\chi\right\}=\\
\displaystyle
\zeta_1\frac{1}{2}\int\,d\tau\,\left\{ -\dot{\phi}\dot{\chi}-F\dot{\chi}-\ddot{\phi}\chi-\dot{F}\chi\right\}=-\zeta_1\frac{1}{2}\int\,d\tau\,\frac{d}{d\tau}\left\{\left(\dot{\phi}+F\right)\chi\right\}
\end{array}
\end{equation}
Its contribution to the second variation is given by the expression
\begin{equation}
\label{delta2Sfree_boundary}
\begin{array}{l}
\displaystyle
\delta_2 S_\mathrm{free}^{(b)}=\int\,d\tau\,\frac{1}{2}\left(\psi\delta_2\dot{\chi}+\dot{\psi}\delta_2\chi\right)=
-\zeta_2\frac{1}{2}\int\,d\tau\,\left\{\psi(\ddot{\phi}-\dot{F})+\dot{\psi}(\dot{\phi}-F)\right\}=\\
\hskip1.5truecm
\displaystyle
-\zeta_2\frac{1}{2}\int\,d\tau\,\frac{d}{d\tau}\left\{\left(\dot{\phi}-F\right)\psi\right\}
\end{array}
\end{equation}
Assembling all the terms we find the following expressions for the two variations:
\begin{equation}
\label{supercharges1}
\begin{array}{l}
\displaystyle
\delta_1 S_\mathrm{free}= -\zeta_1\frac{1}{2}\int\,d\tau\,\frac{d}{d\tau}\left\{\left(\dot{\phi}+F\right)\chi\right\}\\
\displaystyle
\delta_1 S_\mathrm{int} =\zeta_1\int\,d\tau\,\frac{d}{d\tau}\left(W'\chi\right)\\
\displaystyle 
\delta_2 S_\mathrm{free}=\zeta_2\frac{1}{2}\int\,d\tau\,\frac{d}{d\tau}\left\{\left(\dot{\phi}-F\right)\psi\right\}\\
\displaystyle
\delta_2 S_\mathrm{int}=\zeta_2\int\,d\tau\,\frac{d}{d\tau}\left(W'\psi\right)
\end{array}
\end{equation}
Therefore the conserved charges are given by the expressions, 
\begin{equation}
\label{conserved_charges}
\begin{array}{l}
\displaystyle
\bm{Q}_1=\frac{1}{2}\left(-\dot{\phi}-F+2W'(\phi)\right)\chi=\frac{1}{2}\left(-\dot{\phi}+W'(\phi)\right)\chi\\
\displaystyle
\bm{Q}_2=\frac{1}{2}\left(\dot{\phi}-F+2W'(\phi\right)\psi = \frac{1}{2}\left(\dot{\phi}+W'(\phi)\right)\psi
\end{array}
\end{equation}
once we replace $F=W'(\phi)$, which we may do, since $F$ enters quadratically in the action.

In matrix notation, defining $\bm{Q}_\alpha\equiv(\bm{Q}_1,\bm{Q}_2)^\mathrm{T}$ and recalling that $\Psi_\alpha\equiv(\psi,\chi)^\mathrm{T}$,
\begin{equation}
\label{conserved_charges_1}
\bm{Q}_\alpha = \frac{1}{2}\left(-\varepsilon_{\alpha\beta}\dot{\phi} + \left[\sigma_1\right]_{\alpha\beta}W'(\phi)\right)\Psi_\beta
\end{equation}
It remains to show that these charges do realize a representation of the supersymmetry algebra. As usual, this can be achieved, either in the Hamiltonian formalism, or in the Lagrangian formalism. The Hamiltonian formalism is presented in all the textbooks on the subject~\cite{Wess:1992cp,Freund:1986ws,Argyres:2001eva,Gates:1983nr,West:1990tg}, where the crucial point is that the two anticommuting fields, $\psi(\tau)$ and $\chi(\tau)$ are canonically conjugate and, therefore, their anticommutation relations are $\{\psi(\tau),\psi(\tau')\}=\delta(\tau-\tau')=\{\chi(\tau),\chi(\tau')\}$ but that $\{\psi(\tau),\chi(\tau')\}=0$.

 It is the ``deformation'' expressed by the first two relations, with respect to the ``classical'' relations, that leads to the conclusion that the anticommutator of the two charges is non--zero. 
 
This, however, isn't enough. The subtlety is that the anticommutator of the charges does give rise to a contribution proportional to the Hamiltonian, only upon analytic continuation to real time--or upon analytic continuation of the auxiliary field. And the reason behind that is that the equation $F=W'(\phi)$, the equation of motion for the auxiliary field, in fact, can only make sense as the pair of equations $F=0=W'(\phi)$, since the auxiliary field takes purely imaginary values, while the scalar takes real values. 

On the other hand, if, in the final expressions~(\ref{conserved_charges}) or (\ref{conserved_charges_1}) we, either, rotate $\tau\to\mathrm{i}t$,  or we set 
$W'(\phi)\to\mathrm{i}W'(\phi)$, in Euclidian signature, then we do, indeed, find that 
\begin{equation}
\label{charges_algebra}
\left\{\bm{Q}_\alpha,\bm{Q}_\beta\right\}=-\left(\dot{\phi}^2+W'(\phi)^2\right)\delta_{\alpha\beta}=-2H\delta_{\alpha\beta}
\end{equation}

This, naturally, highlights that the analytic continuation from Euclidian to Minkowski signature is quite subtle~\cite{Zumino:1977yh} and has, correspondingly,  non--trivial implications for the construction of superspace, that deserve a fuller discussion (that may be found in the book~\cite{West:1990tg} in the chapters dealing with Euclidean superspace). 
 Since we shall not use the superspace formulation in what follows, we shall not pursue the analysis here.

 Let us, however, recall that the reason this formulation is useful is that, at the level of the classical action, at least, it implies that the supersymmetry transformations in terms of components~(\ref{SUSYalgebra}) can be written as transformations in superspace induced by transformations on $(\tau,\theta,\overline{\theta})\to(\tau+z_\alpha\varepsilon^{\alpha\beta}\Theta_\beta,\theta+\zeta_1,\overline{\theta}+\zeta_2)$, where $z_\alpha\equiv(\zeta_1,\zeta_2)$, and $\Theta_\alpha\equiv(\theta,\overline{\theta})$, as has been stressed in the standard textbooks on the subject~\cite{Wess:1992cp,Freund:1986ws,Argyres:2001eva,Gates:1983nr,West:1990tg}.
 
It is the property that these transformations act in this way, that implies that the measure $d\tau\,d\theta\,d\overline{\theta}$ is invariant--and that the measure of the partition function, $[\mathscr{D}\phi(\tau)][\mathscr{D}\psi(\tau)][\mathscr{D}\chi(\tau)]$, assuming that it can be well-defined, e.g. on the lattice,
is, also.  

What is by no means obvious, however, is how one is supposed to ``count'' the points along $\theta$ and $\overline{\theta}$, what the integration measure $d\theta\, d\overline{\theta}$ actually measures. 

A hint can be found in the fact that the anticommuting degrees of freedom represent the determinant of a (differential) operator in a local way in the action; therefore what they're counting is its rank. 

We shall show in the subsequent sections how it is possible to evade having to deal with this issue on the lattice.  We shall explore how supersymmetry, that implies the existence of supermultiplets,  that include commuting and anticommuting fields, in a way that allows their mixing, will allow us to describe the effects of the anticommuting fields using only commuting fields.   What is the important point is that the position $\phi(\tau)$ can be consistently interpreted as a component of  superfields  and, in the same way that these structures have physical meaning, while their individual components do not, it is these structures that describe the physics, when fluctuations are relevant--that's the bottom line. And in the following sections we shall show what this statement means, in practice: that it is the identities satisfied by the noise fields, functionals of the scalar, that ensure the system is consistently closed.
 
Specifically, inasmuch as the correlation functions of the noise field, sampled with $Z_\mathrm{QM}$, are found to satisfy Wick's theorem, this means that the fluctuations induce the presence of the absolute value of the stochastic determinant. Of course, from the correlation functions of the noise field, it is possible to deduce the correlation functions of the scalar and of the anticommuting fields, by acting with the supercharges. The practical details of the dictionary, must be worked out.

The conclusion here is that the classical action, obtained from the Langevin equation, and including the Jacobian of the transformation from the noise to the scalar, is, indeed, invariant under supersymmetry transformations--whatever the properties of the, classical, (super)potential. This raises, as mentioned above, two questions: do fluctuations affect this invariance; and what is  the appropriate classical action, that can be used within the usual path--integral formalism? The first question has been extensively studied~\cite{Kastner:2007gz,Wozar:2011gu,Schierenberg:2012pb,Bergner:2009vg,Bergner:2012nu,Baumgartner:2015zna,Baumgartner:2015qba,Baumgartner:2014nka,Baumgartner:2013ara,Baumgartner:2011jw,Baumgartner:2011cm,Baumgartner:2012np,Bergner:2016qbz,Bergner:2015ywa}; the second, much less, beyond
~\cite{parisi_sourlas},~\cite{deAlfaro:1984gw,deAlfaro:1984hb,DeAlfaro:1986uv} and
~\cite{Cooper:1982dm,Bender:1983xi,Freedman:1983as} and, even then, it wasn't the focus of their attention.   It has been always assumed that supersymmetry required a classical action that contained the superpartners explicitly, that they couldn't be generated by the fluctuations and that the index is the only relevant quantity. 

These assumptions are, in fact, incorrect. The stochastic formulation, indeed, shows that the superpartners can incarnate the fluctuations and it has been known for some time that ``wall crossing'' can occur, where the index changes value. What hasn't been appreciated is that it isn't necessary to calculate every single factor, that relates $Z_\mathrm{QM}$ and $Z_\mathrm{L}$ separately and explicitly and that it is possible to use the action of the commuting variables to 
compute the correlation functions of the noise field, since it is, in fact a function only of these. And use the supersymmetry transformations to obtain the relations between these correlation
functions and those of the fermions. 

In addition, beyond any theoretical interest, however, the experimental situation has changed dramatically the last thirty years and experiments that probe fundamental properties of quantum matter have passed from the realm of thought experiments to that of reality. The study of new properties and phases of quantum matter, that has become possible, motivates, also, a fresh look. We shall use lattice techniques to compute the identities that should be satisfied by the correlation functions of the scalar, sampled using $Z_\mathrm{QM}$ to show that ordinary quantum mechanics does define observables that can probe this, unexpected, symmetry.

These results, also, mean that the non--relativistic particle, when fluctuations are taken into account, doesn't probe, simply the space, parametrized by the commuting coordinate $\phi$, but a space that is, also, parametrized by  the anticommuting coordinates $\psi$ and $\chi$ (it's possible to consistently replace the auxiliary field, $F$, by its equation of motion). This is, of course,  superspace, a particular case of a non--commutative space.  There's been much work in non--commutative geometry, both from the mathematical and the physical point of view; however, while it has been recognized that quantum effects render phase space non-commutative (though not all non--commutative phase spaces describe quantum effects), what has received much less attention is the fact that fluctuations in general have as natural framework non--commutative spaces of a particular kind, namely superspaces, where both worldvolume and target space acquire anticommuting dimensions, that mix with the commuting dimensions in a way described by supersymmetry and that the fields that provide the natural description are superfields. Therefore it's possible to acquire intuition about supersymmetry and its role for describing physical effects  by studying it in many more contexts than hitherto acknowledged.

 \section{Beyond the classical action: the correlation functions on the lattice}\label{SUSY_qu}
In the previous sections we have reviewed the properties of the classical action, defined by the partition functions $Z_\mathrm{L}$ and $Z_\mathrm{SUSY}$--to probe the effects of the fluctuations we shall study the identities satisfied by the correlation functions, computed with the partition function $Z_\mathrm{QM}$ and in a way that doesn't depend on any perturbative expansion.This is possible using a lattice regularization. 

\subsection{The lattice setup}\label{lattice}
We introduce an infrared cutoff, $0\leq\tau\leq T$, and an ultraviolet cutoff, $a=T/N$, $\tau\equiv n\cdot a$, $0\leq n\leq N$.  We shall impose periodic boundary conditions on the scalar, 
$\phi(\tau)\equiv\phi_n$, i.e. $\phi_n=\phi_{n+N}$. 

In this way the ill--defined measure, $[{\mathscr D}\phi]$, becomes quite explicitly defined:
\begin{equation}
\label{field_measure}
[{\mathscr D}\phi]\equiv\prod_{n=0}^{N-1}\,d\phi_n
\end{equation}

Replacing the derivatives by finite differences, we write the scalar action (having 
dropped the total derivative, thanks to the boundary conditions)
\begin{equation}
\label{Sscalar}
S[\phi]=\int\,d\tau\,\left[ \frac{1}{2}\left(\frac{\partial\phi}{\partial\tau}+\frac{\partial W}{\partial\phi}\right)^2\right]= 
\int\,d\tau\,\left[ -\frac{1}{2}\phi\frac{\partial^2}{\partial\tau^2}\phi + \frac{1}{2}\left(\frac{\partial W}{\partial\phi}\right)^2\right]
\end{equation}
as follows:
\begin{equation}
\label{Sscalarlatt}
S[\phi] = a\sum_{n=0}^{N-1}\left[ 
-\phi_n\frac{\phi_{n+1}-2\phi_n+\phi_{n-1}}{2a^2}+\frac{1}{2}\left(\frac{\partial W}{\partial\phi_n}\right)^2
\right]
\end{equation}
The superpotential, $W(\phi_n)$, we shall consider will be the polynomial superpotential,
\begin{equation}
\label{quarticW}
W(\phi_n) = c\phi_n+\frac{m^2}{2}\phi_n^2 + \frac{\lambda}{K!}\phi_n^K 
\end{equation}
In this way we can treat the two special cases, $K=3$ and $K=4$ together and will try and understand what happens, when $W''(\phi)$ is, nonetheless, of fixed sign. For both cases we remark that the scalar potential is bounded from below and confines at infinity, therefore a ground state always exists. However, while, in both cases, by choosing the parameters appropriately, the classical minimum is unique, nonetheless, in the presence of fluctuations, the two potentials describe different limiting cases.  

The linear term doesn't play a significant role for the quartic superpotential: since $W'(\phi)$ is a cubic polynomial, with real coefficients, therefore the equation $V'(\phi)=0=(W'(\phi)) W''(\phi)$ always possesses a real root, so a supersymmetric vacuum always exists, if $W''(\phi)\neq 0$, as assumed.  The only effect of the linear term is, whether there are three or one real roots. The quantum (super)potential, of course, possesses only one minimum. 

On the other hand, the linear term  is relevant for the classical cubic superpotential. The reason is that it is this term that controls whether the classical scalar potential possesses one or two minima:
\begin{equation}
\label{cubic_superp_scalar_pot}
V(\phi)=\frac{1}{2}W'^2(\phi)=\frac{1}{2}\left(c + m^2\phi + \frac{\lambda}{2}\phi^2\right)^2 = 
\frac{1}{2}\left( \frac{\lambda}{2}\left(\phi+\frac{m^2}{\lambda}\right)^2 + c-\frac{m^4}{2\lambda}\right)^2
\end{equation}
We deduce readily that, for $c<(m^4/(2\lambda))$, the (classical) scalar potential possesses two, degenerate minima, while, for $c\geq (m^4/(2\lambda))$, it possesses only one. If $c\leq (m^4/(2\lambda))$,  at these minima, the scalar potential vanishes, therefore, supersymmetry may be realized, the question is, in which mode, Wigner or Nambu--Goldstone. For $c>(m^4/(2\lambda))$, the scalar potential possesses a single minimum, where it does not vanish--supersymmetry is broken explicitly. So the question really is, what are the properties of the quantum (super)potential. These will be probed by the identities of the correlation functions.

Let us define the lattice parameters
\begin{equation}
\label{lattparam}
\begin{array}{cccc}
\displaystyle \varphi_n\equiv a^{-1/2}\phi_n, & \displaystyle m_\mathrm{latt}^2\equiv m^2 a, & \displaystyle \lambda_\mathrm{latt}\equiv \lambda a^{\frac{K}{2}}, & \displaystyle c_\mathrm{latt}\equiv c a^{1/2}
\end{array}
\end{equation}
in terms of which the lattice action becomes
\begin{equation}
\label{Slattscaled}
S_\mathrm{latt}[\varphi]=\sum_{n=0}^{N-1}\left[-\varphi_n\varphi_{n+1}+\varphi_n^2+\frac{1}{2}\left(c_\mathrm{latt}+m_\mathrm{latt}^2\varphi_n+\frac{\lambda_\mathrm{latt}}{(K-1)!}\varphi_n^{K-1}\right)^2\right]
\end{equation}
In this expression the lattice spacing is no longer present--it's ``hidden'' in the lattice parameters, $c_\mathrm{latt},m_\mathrm{latt}$ and $\lambda_\mathrm{latt}$.  To define the scaling limit, it's useful to work with quantities that have well--defined scaling properties.

In this expression we have used a different convention than is customary in the literature, since we haven't absorbed the $\varphi_n^2$ term, that comes from the discretization of the kinetic term, into a redefinition of the mass on the lattice. This is for convenience, of course, in writing the action  as the sum of squares, up to surface terms, that are absent, due to the periodic boundary conditions and terms that are proportional to positive powers of the lattice spacing.
 
We, also, remark that 
\begin{equation}
\label{effective_coupling}
g\equiv\frac{\lambda}{|m|^{K}}=\frac{\lambda_\mathrm{latt}}{|m_\mathrm{latt}|^{K}}
\end{equation}
This combination, which vanishes for vanishing coupling constant, therefore can be taken as an effective measure of its strength, is (classically) scale invariant. Similarly, we remark that the ratio
\begin{equation}
\label{effective_boundary_term}
\frac{c_\mathrm{latt}}{|m_\mathrm{latt}|}=\frac{c}{|m|}\equiv C
\end{equation}
also doesn't seems to depend on the scale, defined by the lattice spacing, while 
\begin{equation}
\label{effective_scale}
s\equiv\frac{|m^2|}{\lambda}= a^{\frac{2-K}{2}}\frac{|m_\mathrm{latt}|^2}{\lambda_\mathrm{latt}}
\end{equation}
provides a good definition of a length scale, for non--quadratic potentials, for which $K\neq 2$.  The scaling limit may thus be defined by taking the ratio 
$m_\mathrm{latt}^2/\lambda_\mathrm{latt}\to 0$, as the lattice spacing $a\to 0$ and the lattice size, $N\to\infty$, keeping the ratio $m^2/\lambda$  fixed. The ratio $g\equiv\lambda/|m|^{K}$ is arbitrary--if it's less than 1, the theory is weakly coupled, if it's greater than, or of order 1, the theory is strongly coupled.  We remark that the case $s=0$ requires special care and we defer its study.

 In what follows we shall present first results of the numerical treatment, where we study the effect of reducing $|m_\mathrm{latt}|$, as we increase the system  size, $N$, in order to test the proximity to the scaling limit.

The partition function, at finite lattice spacing and lattice size, thus,  takes the following form
\begin{equation}
\label{Zlatt}
Z_\mathrm{latt}=\int\,\left[\prod_{n=0}^{N-1} d\varphi_n\right]\,
e^{-\sum\left[-\varphi_n\varphi_{n+1}+\varphi_n^2+\frac{1}{2}\left(c_\mathrm{latt}+m_\mathrm{latt}^2\varphi_n+\frac{\lambda_\mathrm{latt}}{(K-1)!}\varphi_n^{K-1}\right)^2\right]}
\end{equation}
We would like to show that, up to terms that vanish in the limit of vanishing lattice spacing, the expression in the exponent is a perfect square, namely, 
\begin{equation}
\label{auxFlatt2}
\begin{array}{l}
\displaystyle 
-\varphi_n\varphi_{n+1}+\varphi_n^2+\frac{1}{2}\left(c_\mathrm{latt}+m_\mathrm{latt}^2\varphi_n+\frac{\lambda_\mathrm{latt}}{(K-1)!}\varphi_n^{K-1}\right)^2 = \\
\displaystyle
\frac{1}{2}\left(   \left(\varphi_{n+1}-\varphi_n\right)^2 + \left(c_\mathrm{latt}+m_\mathrm{latt}^2\varphi_n+\frac{\lambda_\mathrm{latt}}{(K-1)!}\varphi_n^{K-1}\right)^2 \right)=\\
\displaystyle
\frac{1}{2}\left( 
\varphi_{n+1}-\varphi_n +c_\mathrm{latt}+ m_\mathrm{latt}^2\varphi_n+\frac{\lambda_\mathrm{latt}}{(K-1)!}\varphi_n^{K-1}
\right)^2-\left(\varphi_{n+1}-\varphi_n\right) W'(\varphi_n;c_\mathrm{latt},m_\mathrm{latt}^2,\lambda_\mathrm{latt})
\end{array}
\end{equation}
If we expand the last term in powers of the lattice spacing we find that
\begin{equation}
\label{totderauxFlatt}
\begin{array}{l}
\displaystyle
\left(\varphi_{n+1}-\varphi_n\right) W'(\varphi_n;c_\mathrm{latt},m_\mathrm{latt}^2,\lambda_\mathrm{latt}) = \\
\displaystyle
a\frac{d}{d\tau}W(\varphi_n;c_\mathrm{latt},m_\mathrm{latt}^2,\lambda_\mathrm{latt}) + 
\sum_{m=1}^\infty\frac{a^{2m+1}}{(2m+1)!}
\left[\frac{d}{d\tau}\left(\frac{d^{2m}\varphi}{d\tau^{2m}}W'\right)-\frac{d^{2m}\varphi}{d\tau^{2m}}\frac{d}{d\tau}W'\right]
\end{array}
\end{equation}
The superpotential, $W(\varphi_n;c_\mathrm{latt},m_\mathrm{latt}^2,\lambda_\mathrm{latt})$, has a smooth limit, as $a\to 0$, so does not cancel the explicit lattice spacing factor:
\begin{equation}
\label{Wlattcont}
W(\varphi_n;c_\mathrm{latt},m_\mathrm{latt}^2,\lambda_\mathrm{latt})=(ca^{1/2})a^{-1/2}\phi_n + \frac{1}{2}(m^2a)a^{-1}\phi_n^2 + \frac{1}{K!}(\lambda a^{K/2})a^{-K/2}\phi_n^K=W(\phi_n,m^2,\lambda)
\end{equation}
We remark that the $O(a)$ term is, apparently, a surface term and the higher order terms are a sum of surface terms and terms that vanish in the limit of vanishing lattice spacing, provided the series converges. Assuming that their contribution in correlation functions is not cancelled by a linear divergence, they may, thus, be neglected, in the continuum limit. Their presence expresses how supersymmetry is broken by the lattice regularization. 

This can be understood by realizing that, up to the lattice artifacts, identified above, the partition function takes the form
\begin{equation}
\label{Zlatt1}
\begin{array}{l}
\displaystyle
Z_\mathrm{latt}=\int\,\prod_{n=0}^{N-1}\,d\varphi_n\,e^{\sum_{n=0}^{N-1}\,\frac{1}{2}\left( \varphi_{n+1}-\varphi_n +c_\mathrm{latt}+ m_\mathrm{latt}^2\varphi_n+\frac{\lambda_\mathrm{latt}}{(K-1)!}\varphi_n^{K-1} \right)^2}=\\
\displaystyle
\hskip1truecm
\prod_{n=0}^{N-1}\,\int\,d\varphi_n\,e^{\frac{1}{2}\left( \varphi_{n+1}-\varphi_n +c_\mathrm{latt}+ m_\mathrm{latt}^2\varphi_n+\frac{\lambda_\mathrm{latt}}{(K-1)!}\varphi_n^{K-1} \right)^2}
\end{array}
\end{equation} 
What's missing for this expression to be equal to 1, is the factor $|\mathrm{det}\,{\sf A}|^{-1}$, where ${\sf A}_{n,n'}=\partial \eta_n/\partial\varphi_{n'}$, with
\begin{equation}
\label{noiselatt1}
\eta_n\equiv \varphi_{n+1}-\varphi_n +c_\mathrm{latt}+ m_\mathrm{latt}^2\varphi_n+\frac{\lambda_\mathrm{latt}}{(K-1)!}\varphi_n^{K-1}
\end{equation}
Were the term $|\mathrm{det}\,{\sf A}|^{-1}$ present, the lattice regularization would have realized supersymmetry. The question is, whether the fluctuations can generate it. 

The lattice calculation of the correlation functions will allow us to test these assumptions.  It might, therefore, be useful to investigate whether ``perfect lattice actions''~\cite{Hasenfratz:1993sp} could be written for this case, that could reduce the systematic errors. (Of course these do not eliminate possible non-analytic dependence on the lattice spacing. But such singularities reflect physical effects, like tunneling, or phase transitions, that herald the appearance of new degrees of freedom.)

This expression, for the lattice partition function, corresponds to the lattice version of $Z_\mathrm{QM}$, that neglects the lattice artifacts; and we would like to check that, if we sample the space of field configurations with the lattice action that is not a sum of squares, we do obtain results consistent with the absence of the artifacts. What this calculation allowed us to accomplish is to identify the lattice counterpart of the noise field. 

It should be stressed at this point, that eq.~(\ref{noiselatt1}) implies that it isn't possible to distinguish, unambiguously, the value of $c_\mathrm{latt}$, the coefficient of the linear term of the superpotential, from the value of the expectation value of the noise field, $\langle\eta_n\rangle$. Indeed we realize that it is possible to tune the value of $c_\mathrm{latt}$, so that $\langle\eta_n\rangle=0$--which means that supersymmetry won't be broken. However, if it turns out that the reason it isn't is because $c_\mathrm{latt}\neq 0$, this means that it is boundary terms that are responsible for preserving it; whereas, if  $c_\mathrm{latt}=0$, boundary terms don't contribute. 

It is useful to perform one more change of variables, in order to simplify the expressions in view of the numerical treatment. 
Let us define the field $\bm{\varphi}_n$
\begin{equation}
\label{newPhi}
\varphi_n\equiv\left(\frac{|m_\mathrm{latt}|^2}{\lambda_{\mathrm{latt}}}\right)^\alpha\bm{\varphi}_n
\end{equation}
By choosing $\alpha\equiv 1/(K-2)$, the expression for the lattice action simplifies considerably: 
\begin{equation}
\label{Slattrescaled}
\begin{array}{l}
\hskip-0.5truecm
S_\mathrm{latt}=\frac{1}{g^{\frac{2}{K-2}}m_\mathrm{latt}^2}\sum_{n=0}^{N-1}
\left\{ -\bm{\varphi}_n\bm{\varphi}_{n+1}+\bm{\varphi}_n^2+\frac{m_\mathrm{latt}^4}{2}\left( Cg^{\frac{1}{K-2}}+\bm{\varphi}_n+\frac{\bm{\varphi}_n^{K-1}}{(K-1)!}\right)^2\right\}
\\
\end{array}
\end{equation} 
where we have, also, used the definition of the dimensionless coupling constant, $g$. It is this expression, in fact, that's most useful for the Metropolis algorithm. We remark that it, also, allows to treat equally easily the case of a unique and that of multiple classical minima. The details of the latter case will be presented in a separate publication, in order not to overload the presentation here.

We remark that eq.~(\ref{Slattrescaled}) implies that the partition function, $Z_\mathrm{QM}$, can be written as
\begin{equation}
\label{ZlattQM}
Z_\mathrm{QM}=\int\,\prod_{n=0}^{N-1}\,\left[d\bm{\varphi}_n\right]\,e^{-\frac{1}{2g^{2/(K-2}|m_\mathrm{latt}^2|}\sum_{n=0}^{N-1}\bm{\eta}_n^2}
\end{equation}
This implies  that, if the fluctuations do produce the absolute value of the Jacobian $\partial\varphi_n/\partial\eta_{n'}$, then 
\begin{equation}
\label{2ptnoise}
\left\langle\bm{\eta}_n\bm{\eta}_{n'}\right\rangle=g^{\frac{2}{K-2}}|m_\mathrm{latt}^2|\delta_{|n-n'|,0}
\end{equation}
which is another, quite concrete, test of the numerical simulations. 

It should be noted that, when we impose periodic boundary conditions, we are computing 
{\em half} this value, as we shall see in the discussion of the numerical simulations. 

In the following we shall describe in more detail the observables and their expected behavior, when the classical potential has a unique minimum. To simplify calculations we shall study the cubic and quartic superpotentials, that are representative of  two classes of behavior and discuss how to reveal the subtleties that are hidden, when the cubic superpotential can, nevertheless, give rise to a scalar potential that possesses a unique minimum. 

But, before doing that, it's useful to understand that the lattice formulation presented here, is, indeed, equivalent, up to redefinition of the parameters, to the ``usual'' lattice setup, that we recall in the following subsection. It is in this way that we shall understand that supersymmetry is an inevitable--though well hidden--property of ``usual'' descriptions of a particle, in equilibrium with a bath. 

\subsection{The usual lattice setup}\label{usuallatt}
The lattice setup presented in the previous section, has the advantage that it shows how supersymmetry might be detected among the artifacts; however it suffers from the suspicion that it was invented for that purpose. After all, lattice studies of anharmonic oscillators have been done for decades, so the lattice artifacts are under control; it was, just, that the observables, that correspond to the noise field were, up to now, not studied. So to really show that supersymmetry is an inevitable property of the dynamics of a particle, in equilibrium with a bath, we must show how, using the ``usual'' lattice action, but measuring the correlation functions of the noise field, it is, nevertheless, possible to find that these correlation functions do define a Gaussian field, with ultra--local 2--point function. 

In the usual approach, one works with the scalar potential, not the superpotential:
\begin{equation}
\label{scalarpot4}
V_4^{(0)}(\phi_0)=c_0+\frac{m_0^2\phi_0^2}{2}+\frac{\lambda_4}{4!}\phi_0^4
\end{equation}
for the case of the quartic oscillator  and 
\begin{equation}
\label{scalarpot6}
V_6^{(0)}(\phi_0)=c_0+\frac{m_0^2\phi_0^2}{2}+\frac{\lambda_4}{4!}\phi_0^4+\frac{\lambda_6}{6!}\phi_0^6
\end{equation}
for the case of the sextic oscillator.

What the knowledge about the existence of the superpotential implies is that these expressions can be written as perfect squares, up to additive constants, that, while depending on the dynamical variables, don't contribute to the correlation functions. What isn't obvious, however, is, whether 
the perfect squares are, also, polynomials.

For $V_4^{(0)}(\phi_0)$, we readily find that it can, indeed, be written as 
\begin{equation}
\label{V4scalarsquare}
V_4^{(0)}(\phi_0)=\left[\sqrt{\frac{\lambda_4}{4!}}\left(\phi_0^2+\frac{6m_0^2}{\lambda_4}\right)\right]^2 + c_0-\frac{3m_0^4}{2\lambda_4}
\end{equation}
which allows us to identify the corresponding superpotential, $W_4^{(0)}(\phi_0)$ as
\begin{equation}
\label{W4scalar0}
\frac{dW_4^{(0)}}{d\phi_0}=\sqrt{2}\sqrt{\frac{\lambda_4}{4!}}\left(\phi_0^2+\frac{6m_0^2}{\lambda_4}\right)\Leftrightarrow
W_4^{(0)}(\phi_0) = \frac{1}{3}\sqrt{\frac{\lambda_4}{4!}}\phi_0^3 + \sqrt{\frac{\lambda_4}{12}}\frac{6m_0^2}{\lambda_4}\phi_0+\widetilde{c}_0
\end{equation}
We note that, while it may not look like the most general expression for a cubic superpotential, it, actually, is, since it's well known that any cubic polynomial of one variable can be brought to the form defined by the above expression, by shifting the field $\phi_0$ in order to eliminate the term in $\phi_0^2$. 
The corresponding noise field is given by the expression 
\begin{equation}
\label{noisefield4usual}
\eta(\tau) = \frac{d\phi_0}{d\tau} + \frac{dW_4^{(0)}}{d\phi_0}
\end{equation}
and the continuum action, therefore, is written as
\begin{equation}
\label{Slatt4usual}
S_\mathrm{E}=\int\,d\tau\,\left\{\frac{1}{2}\eta(\tau)^2+c_0-\frac{3m_0^4}{2\lambda_4}\right\}
\end{equation}
which can be discretized according to the prescription presented in the previous section. 

For $V_6^{(0)}(\phi_0)$, on the other hand, it isn't true that the expression in eq.~(\ref{scalarpot6}) can be, always, written, up to an addiitive constant, as the square of a cubic polynomial, whose integral would, therefore, be a quartic polynomial.

A simple calculation shows that it isn't, in general, possible to find $\alpha_0,\beta_0,m$ and $\lambda$ such that the following relation 
\begin{equation}
\label{V6obstr}
V_6^{(0)}(\phi_0)=c_0+\frac{m_0^2\phi_0^2}{2}+\frac{\lambda_4}{4!}\phi_0^4+\frac{\lambda_6}{6!}\phi_0^6=
\beta_0 + \alpha_0\left(m\phi_0 + \frac{\lambda}{6}\phi_0^3\right)^2
\end{equation}
is an identity. In general $V_6^{(0)}(\phi_0)$ can be written as 
\begin{equation}
\label{V62squares}
V_6^{(0)}(\phi_0) = 
\frac{\lambda_6}{6!}\left(\phi_0^3+15\frac{\lambda_4}{\lambda_6}\phi_0\right)^2 +
\phi_0^2
\left(\frac{m_0^2}{2}-\frac{5}{16}\frac{\lambda_4}{\lambda_6}\right)
\end{equation}
We can notice the obstruction in the second term; and remark that, if 
\begin{equation}
\label{finetuningV6}
\frac{m_0^2}{2}-\frac{5}{16}\frac{\lambda_4^2}{\lambda_6}=0
\end{equation}
then the scalar potential is, indeed, the square of a cubic polynomial.

On the other hand, of course, since $V_6^{(0)}(\phi_0)$ is bounded from below, it can, always, be written in the form 
\begin{equation}
\label{V6Wnonpoly}
V_6^{(0)}(\phi_0) = \widetilde{c} + \frac{1}{2}\left(\frac{d\widehat{W}_6^{(0)}}{d\phi_0}\right)^2
\end{equation}
 But $\widehat{W}_6^{(0)}(\phi_0)$ won't be a polynomial in $\phi_0$. However the noise field will, still be defined by the expression 
\begin{equation}
\label{noise6}
\eta(\tau)=\frac{d\phi_0}{d\tau} + \frac{\partial\widehat{W}_6^{(0)}}{\partial\phi_0(\tau)}
\end{equation} 
 and is expected to have the 
 properties of a Gaussian field, with ultra--local 2--point function. 

In the present paper we shall focus on the case where the superpotential itself is a polynomial in the field. 

If one isn't aware of the relevance of supersymmetry, it is, of course, not easy to guess that the correlation functions of these functions of the scalar field could satisfy any particular identities, namely that the 1--point functions vanish, that the 2--point functions are ultra--local, up to lattice artifacts and that connected higher point functions, also, vanish; in particular the Binder cumulant. Nor that the implication of these properties is that the system is consistently closed, through the interaction of the ``apparent'', commuting degree(s) of freedom with ``hidden'', anticommuting, degrees of freedom. 

This is what we shall show in the following sections, by Monte Carlo simulations. The result is that the fluctuations, once all contributions are accounted for, lead to very specific modifications of the scalar potential itself on the one hand; and to the appearance of the absolute value of the determinant of a very specific differential operator, in such a way that $Z_\mathrm{QM}=Z_\mathrm{L}=1$.

\subsection{The classical quartic superpotential with one minimum}\label{uniqueclassmin4}
The prototypical example is the quartic superpotential, with  $m_\mathrm{latt}^2>0$ and $\lambda_\mathrm{latt}>0$. The coefficient of the linear term, $c_\mathrm{latt}$, may have either sign. Under these circumstances there always exists a unique real root to the equation $W'(\phi)=0$, that, of course, does depend on $c_\mathrm{latt}$. For $c_\mathrm{latt}=0$, this root is $\phi=0$. In the present study we shall, therefore, take $c_\mathrm{latt}=0$. 

The observable we shall study is the lattice approximation to the noise term.

In terms of $\bm{\varphi}_n$, the corresponding expression reads
\begin{equation}
\label{latt_noise4}
\bm{\eta}_n=\bm{\varphi}_{n+1}-\bm{\varphi}_n+m_\mathrm{latt}^2\left(Cg^{1/2}+\bm{\varphi}_n + \frac{\bm{\varphi}_n^3}{6}\right)
\end{equation}
It's noteworthy that the noise field displays a non-analytic dependence on the coupling constant, $g$, if the coefficient $C$ of the linear term of the superpotential doesn't vanish. In the present paper this is eliminated, since $C=0$. The subtleties that the linear term of the superpotential gives thus rise to deserve a separate study.  

We remark that, whatever the value of $C$, there always exists a real value of $\varphi_n$, such that $\eta_n=0$. Therefore, we expect that $\langle\eta_n\rangle=0$. Since we have, also, chosen $m_\mathrm{latt}^2>0$,  then $\langle\bm{\varphi}_n\rangle=0$. 

What is not trivially obvious, however, is that $\langle\eta_n\eta_{n'}\rangle=\langle\eta_{|n-n'|}\eta_0\rangle$ (by the periodic boundary conditions) is proprtional to a Kronecker $\delta_{|n-n'|,0}$. 

Let us recall that this relation, also, describes a property of the noise field. So it is a non--trivial property of the fluctuations, as described by the Langevin equation, that, while the 1--point function of the noise field can acquire a non--zero value, the 2--point function of the noise field remains, nevertheless, ultra--local. 

The  partition function in eq.~(\ref{Zlatt1}) isn't obviously that of a Gaussian, since, while  the exponent does factorize over the sites, the measure, apparently, is missing the appropriate Jacobian. Nonetheless, Monte Carlo simulations will show that this correlation function is ultra--local on the lattice. (On the other hand, we expect that the 2--point function of the scalar is not a $\delta-$function.) And we  shall explicitly calculate the Binder cumulants, 
\begin{equation}
\label{Binder_c}
\begin{array}{lcl}
\displaystyle
B(\eta_n)\equiv-\frac{\left\langle\eta_n^4\right\rangle}{3\left\langle\eta_n^2\right\rangle^2}+1 & \mathrm{and} &
\displaystyle
B(\phi_n)\equiv-\frac{\left\langle\phi_n^4\right\rangle}{3\left\langle\phi_n^2\right\rangle^2}+1
\end{array}
\end{equation}
the ultra--local limits of the connected 4--point functions. 

We expect that $B(\eta_n)$ vanishes to numerical precision and $B(\phi_n)$ doesn't vanish. 

\subsection{The classical cubic superpotential with one minimum}\label{uniqueclassical3}
For the (classical) cubic superpotential the linear term plays a crucial role. In terms of the $\bm{\varphi}_n$ field, the noise field has the expression
\begin{equation}
\label{latt_noise_3}
\bm{\eta}_n=\bm{\varphi}_{n+1}-\bm{\varphi}_n+m_\mathrm{latt}^2\left(Cg +\mathrm{sign}(m_\mathrm{latt}^2)\bm{\varphi}_n+\frac{\bm{\varphi}_n^2}{2}\right)
\end{equation}
We observe that, for generic values of $C$, the dependence on the coupling constant $g$ seems analytic; however the mapping between $\bm{\eta}_n$ and $\bm{\varphi}_n$ isn't bijective, since the quadratic equation possesses two roots. 

The exception is when $C=1/(2g)$, when the expression between parentheses is a perfect square and the noise term becomes
\begin{equation}
\label{latt_noise_3a}
\bm{\eta}_n=\bm{\varphi}_{n+1}-\bm{\varphi}_n+\frac{m_\mathrm{latt}^2}{2}\left(\bm{\varphi_n}+\mathrm{sign}(m_\mathrm{latt}^2)\right)^2
\end{equation}
We deduce immediately that, with periodic boundary conditions, we may redefine $\widehat{\bm{\varphi}}_n\equiv\bm{\varphi}_n+\mathrm{sign}(m_\mathrm{latt}^2)$
and find that 
 \begin{equation}
\label{latt_noise_3b}
\bm{\eta}_n=\widehat{\bm{\varphi}}_{n+1}-\widehat{\bm{\varphi}}_n+\frac{m_\mathrm{latt}^2}{2}\widehat{\bm{\varphi}}_n^2
\end{equation}
from which we may expect that $\langle\bm{\eta}_n\rangle\neq 0$; but $\langle\bm{\widehat{\varphi}}_n\rangle=0$ for translation invariant vacua. 

\section{Numerical results}\label{numerics}
We use a standard Metropolis algorithm to generate configurations and compute the correlation functions. 
We tune the parameters so that the acceptance rate is, at least, 70\%.  (If it's too high, the configurations are too correlated, if it's too low they are too few.)
We would like, in principle, to take $m_\mathrm{latt}^2\to 0$, as $N\to\infty$, at fixed $g\equiv\lambda_\mathrm{latt}/m_\mathrm{latt}^4$ and 
fixed scale $s\equiv m^2/\lambda=\lim_{a\to 0}(a( m_\mathrm{latt}^2/\lambda_\mathrm{latt}))$.  In practice, therefore, we shall retain $m_\mathrm{latt}^2$, 
$N$ and $g$ as our parameters.

If we compute the variation, $\delta S[\bm{\varphi}_n]\equiv S[\bm{\varphi}_n+\delta\bm{\varphi}_n]-S[\bm{\varphi}_n]$, of the full lattice action, when we attempt the update $\bm{\varphi}_n\to \bm{\varphi}_n+\delta\bm{\varphi}_n$, we find 
 \begin{equation}
 \label{deltaSMet}
 \begin{array}{l}
 \displaystyle
 \delta S[\bm{\varphi}_n]=\\
 \displaystyle
 \frac{1}{m_\mathrm{latt}^2g^{2/(K-2)}}\left(
 -\delta\bm{\varphi}_n\left(\bm{\varphi}_{n+1}+\bm{\varphi}_{n-1}\right)+\delta\bm{\varphi}_n\left(2\bm{\varphi}_n+\delta\bm{\varphi}_n\right)+
 \frac{m_\mathrm{latt}^4}{2}\left( W'(\bm{\varphi}_n+\delta\bm{\varphi}_n)^2-W'(\bm{\varphi}_n)^2\right)\right)\\
 \end{array}
 \end{equation}
  Here we realize why writing the lattice action in terms of the $\bm{\varphi}_n$ field is useful.  We accept the update if $\exp(-\delta S[\bm{\varphi}_n])>r$, where $r$ is a random number, uniformly distributed in the interval $0\leq r <1$.

\begin{figure}[thp]
 \begin{center}
\subfigure{\includegraphics[scale=0.5]{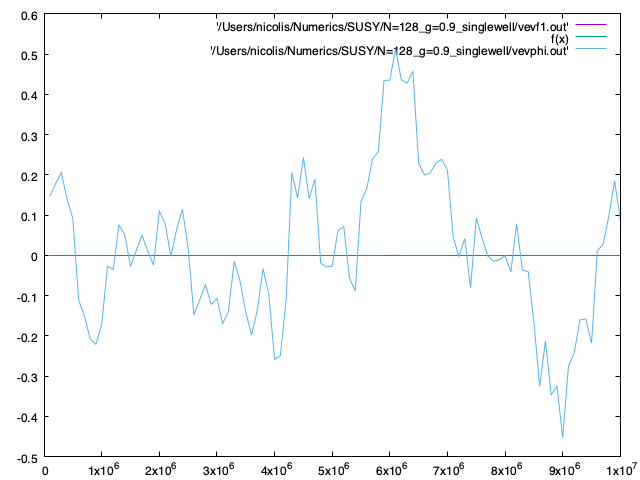}}
\subfigure{\includegraphics[scale=0.5]{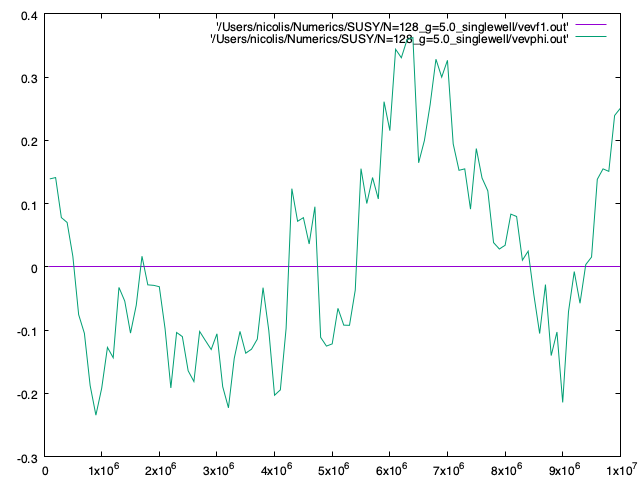}}
 \caption[]{Time series for the 1--point functions, $\langle\bm{\varphi}\rangle$ and $\langle\bm{\eta}\rangle$, for $g=0.9$, and $g=5.0$, $N=128$, $m_\mathrm{latt}^2=10^{-4}$, for the quartic superpotential, with $c=0$.  They imply that  $\langle\bm{\varphi}\rangle=0.0145\pm 0.186$ and $\langle\bm{\eta}\rangle=1.51\times 10^{-6}\pm 1.89\times 10^{-5}$, for $g=0.9$ and that 
 $\langle\bm{\varphi}\rangle=0.013\pm 0.15$ and $\langle\bm{\eta}\rangle=1.37\times 10^{-6}\pm 1.55\times 10^{-5}$, for $g=5.0$,  in  both cases these values are consistent with zero. The fluctuations of the noise field about zero are considerably smaller than those of the scalar.}
 \label{vevfph1_K=4}
 \end{center}
 \end{figure}

 In fig.~\ref{vevfph1_K=4} we display the time series for the 1--point functions, $\langle\bm{\varphi}\rangle$ and $\langle\bm{\eta}\rangle$, for the quartic superpotential, for $g=5.0$ and $m_\mathrm{latt}^2=10^{-4}$, when $N=128$. Both are consistent with zero. 

We compute $\left\langle\bm{\eta}_n\bm{\eta}_{n'}\right\rangle_\mathrm{conn}=\left\langle\bm{\eta}_0\bm{\eta}_{|n-n'|}\right\rangle-\left\langle\bm{\eta}\right\rangle^2$, thanks to translation invariance. This, also, implies that we must take $|n-n'|<(N/2)$. To reduce numerical error of the estimator, it is a good idea to compute, in practice, the ``smeared'' correlator
\begin{equation}
\label{smearedF}
\left\langle\bm{\eta}_{|n-n'|}\bm{\eta}_0\right\rangle = 
\left\langle\bm{\eta}_{|n-n'|+l}\bm{\eta}_l\right\rangle = 
\frac{1}{M}\sum_{m=0}^{M-1}
\frac{1}{N}\sum_{l=0}^{N-1}\bm{\eta}_{\mathrm{mod}\left(|n-n'|+l,N\right)}^{(m)}\bm{\eta}_l^{(m)}
\end{equation}
where $m=0,1,\ldots,M-1$ are the samples. 

 The ultra--local part of the connected 4--point function, 
\begin{equation}
\label{BiC}
B(\bm{\eta})\equiv -\left\langle\bm{\eta}_n^4\right\rangle+3\left\langle\bm{\eta}_n^2\right\rangle^2
\end{equation}
(the so-called {\em Binder cumulant}) should, also, vanish, as a consequence of Wick's theorem, if the noise field is, indeed, drawn from a Gaussian distribution, with $\langle\bm{\eta}\rangle =0$. 
 Of course the full, connected 4--point function should vanish--it's just much easier to check the ultra--local part, which is a necessary condition, anyway. Computationally, it's better to evaluate  the expression
 \begin{equation}
 \label{BiC1}
 B_1(\bm{\eta})\equiv\frac{B(\bm{\eta})}{ 3\left\langle\bm{\eta}^2\right\rangle^2}=
  -\frac{\langle\bm{\eta}^4\rangle}{3\langle\bm{\eta}^2\rangle^2}+1
 \end{equation}
 if we simply want to test whether it's non-zero or not. 
 The reason is that $\langle\bm{\eta}^4\rangle$ and $3\langle\bm{\eta}^2\rangle^2$ are quite comparable in magnitude, so their difference is subject to large numerical uncertainties. Since $3\langle\bm{\eta}^2\rangle^2$ is a strictly positive quantity, the vanishing--or not--of $B$ is equivalent to that of $B_1$. It is, thus, $B_1$ that we will try to estimate.  
 
\begin{figure}[thp]
 \begin{center}
 \subfigure{\includegraphics[scale=0.5]{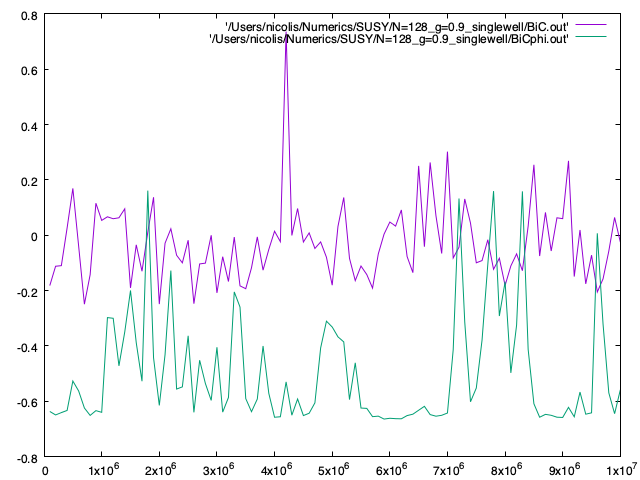}}
 \subfigure{\includegraphics[scale=0.5]{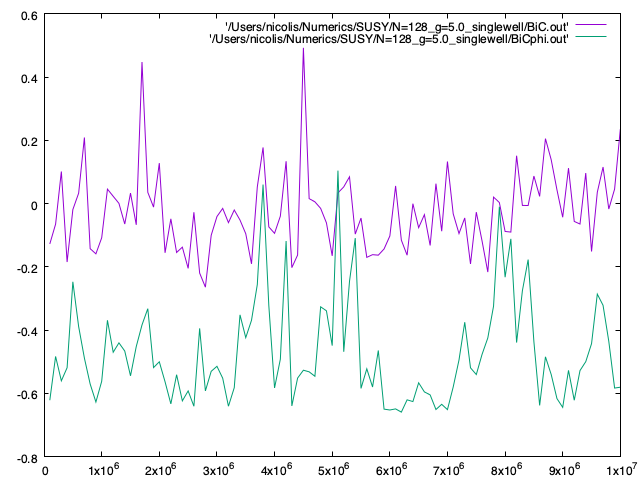}}
 \end{center}
 \caption[]{Time series for the Binder cumulants $-B_1(\bm{\eta})$ and $-B_1(\bm{\varphi})$, for $g=0.9$ and $g=5.0$. They imply that $-B_1(\bm{\eta})=-0.28\pm 0.14$--i.e. consistent with zero--while $-B_1(\bm{\varphi})=-0.49677\pm 0.2018$--i.e. consistent with a non--zero value for $g=0.9$ and $-B_1(\bm{\eta})=-0.026\pm 	0.13$, while 
 $-B_1(\bm{\varphi})=-0.47\pm0.16$ for $g=5.0$.
 }
 \label{BiCtimeseries}
 \end{figure}

 We display the time series for $B_1(\bm{\eta})$ and $B_1(\bm{\varphi})$ in fig.~\ref{BiCtimeseries}. 

In section~\ref{lattice} we showed that these results for the auxiliary field test that the terms of the effective action, when written as a functional of the auxiliary field, that are proportional to positive powers of the lattice spacing, are, truly, irrelevant. 
 
 We hope to improve on these results using Hybrid Monte Carlo methods--especially for the cases where the classical potential possesses multiple minima. 

\begin{figure}[thp]
\begin{center}
\subfigure{\includegraphics[scale=0.7]{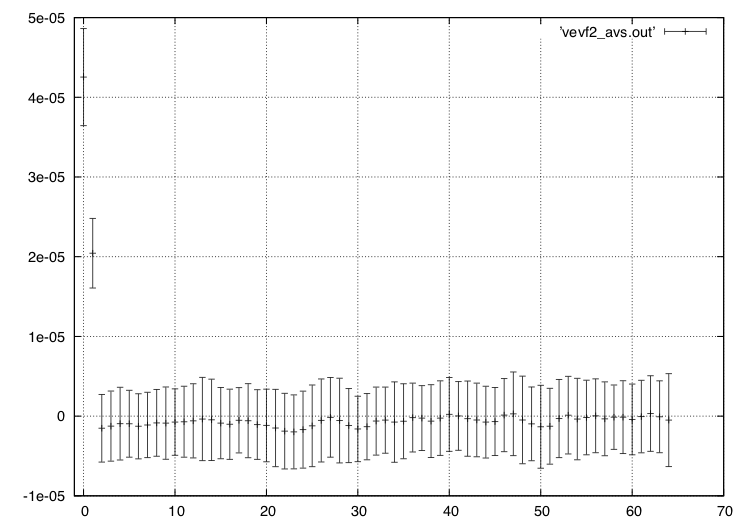}}
\subfigure{\includegraphics[scale=0.5]{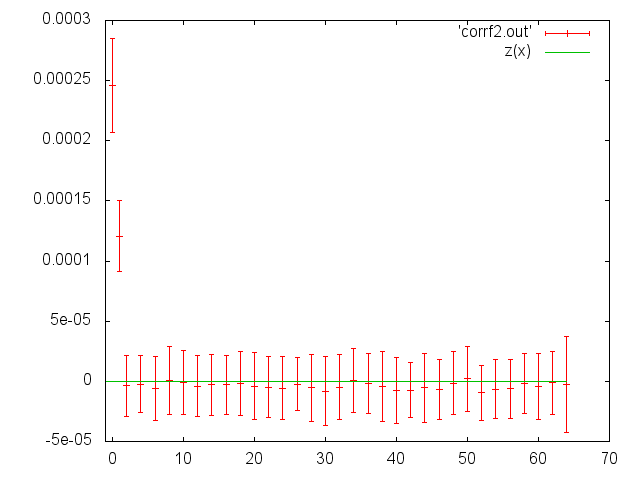}}
\end{center}
\caption[]{The connected 2--point function for the noise field, $\langle\bm{\eta}_{|n-n'|}\bm{\eta}_0\rangle-\langle\bm{\eta}\rangle^2$  for $N=128$, $g=0.9$ and $g=5.0$.The value of $m_\mathrm{latt}^2=0.0001$. We notice that its value, at $|n-n'|=0$, is, indeed, equal to $g|m_\mathrm{latt}^2|/2$, in both cases. }
\label{NumRes128}
\end{figure}

In fig.~\ref{NumRes128}  we display typical plots for the connected 2--point function of the noise field and thus provide support for the thesis that the former is ultra--local,  for $N=128$ and $g=0.9$ (weak coupling), as well as for $g=5.0$ (strong coupling).  A detailed finite size scaling analysis is a priority. 
As expected, the qualitative behavior does not change and supersymmetry is realized in the ``Wigner mode'' at both ``weak'', $g<1.0$  and ``strong'', $g>1.0$ coupling.

What is {\em not} expected is that these results were obtained not by using a manifestly supersymmetric classical action, but using the classical action of the commuting field only. The non--trivial statement is that the corrections, due to the fluctuations, induce in $Z_\mathrm{QM}$ the appearance of the term $|\mathrm{det}(d/d\tau + W''(\phi))|$. 

\begin{figure}[thp]
\begin{center}
\subfigure{\includegraphics[scale=0.4]{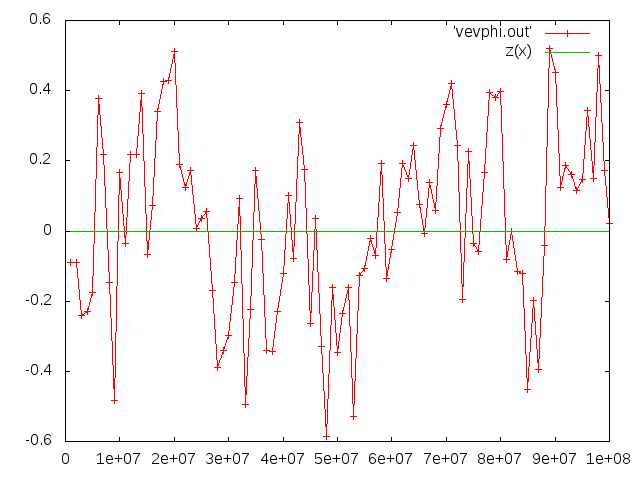}}
\subfigure{\includegraphics[scale=0.4]{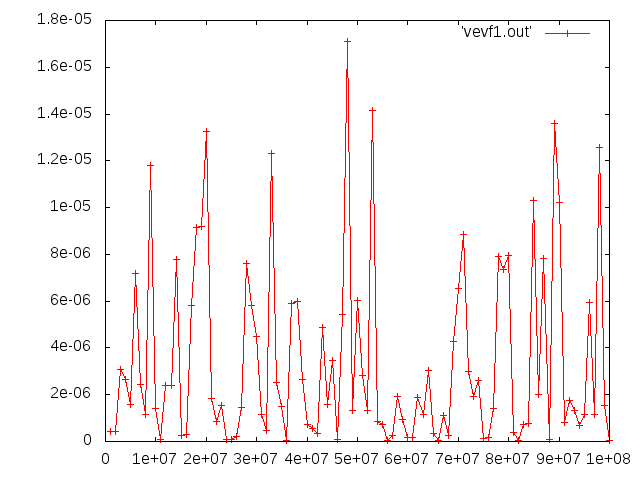}}
\end{center}
\caption[]{The 1--point functions, $\langle\bm{\phi}\rangle$ and $\langle\bm{\eta}\rangle$, for the cubic superpotential, when the scalar potential has a unique minimum, when $g=0.9$. We remark that supersymmetry is spontaneously broken--but $\langle\bm{\eta}\rangle$ is, numerically, very small.}
\label{1ptcubicw}
\end{figure}

\begin{figure}[thp]
\begin{center}
\subfigure{\includegraphics[scale=0.4]{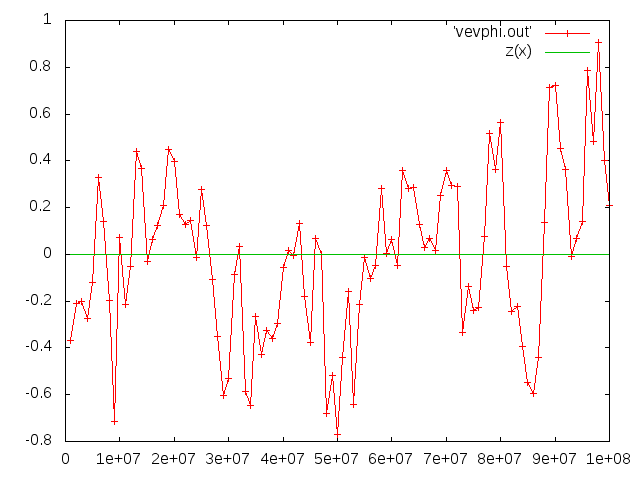}}
\subfigure{\includegraphics[scale=0.4]{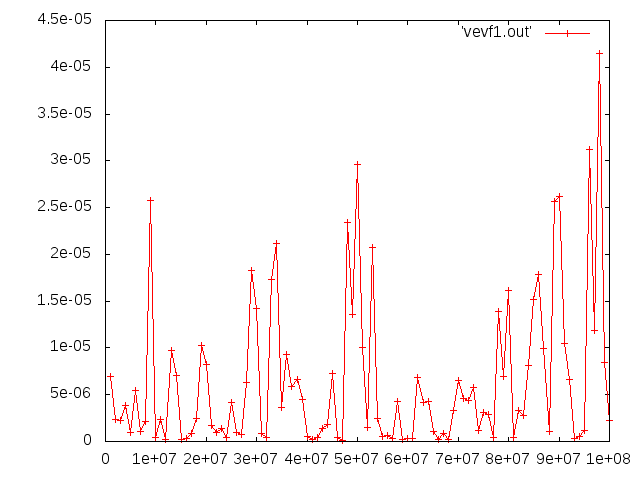}}
\end{center}
\caption[]{The 1--point functions, $\langle\bm{\phi}\rangle$ and $\langle\bm{\eta}\rangle$, for the cubic superpotential, when the scalar potential has a unique minimum, when $g=2.0$. We remark that supersymmetry is spontaneously broken--but $\langle\bm{\eta}\rangle$ is, numerically, very small.}
\label{1ptcubics}
\end{figure}

\begin{figure}[thp]
\includegraphics[scale=0.7]{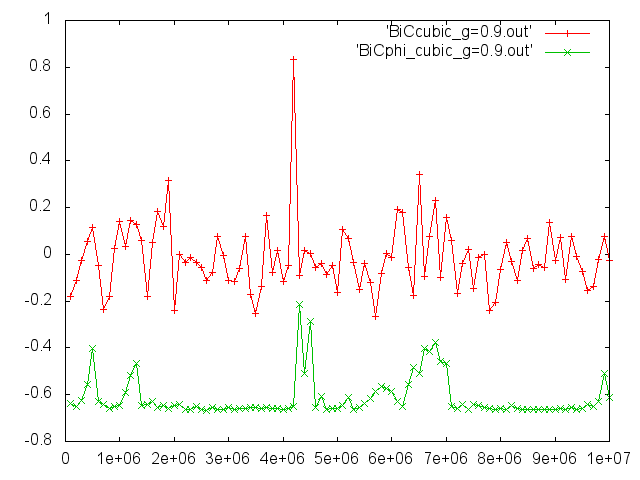}
\caption[]{The time series for the Binder cumulants, $-B_1(\bm{\eta})$ and $-B_1(\bm{\varphi})$, for the cubic superpotential, when $g=0.9$  and the classical scalar potential has a unique minimum. $m_\mathrm{latt}=0.0001$.} 
\label{BiCcubicw}
\end{figure}

\begin{figure}[thp]
\subfigure{\includegraphics[scale=0.5]{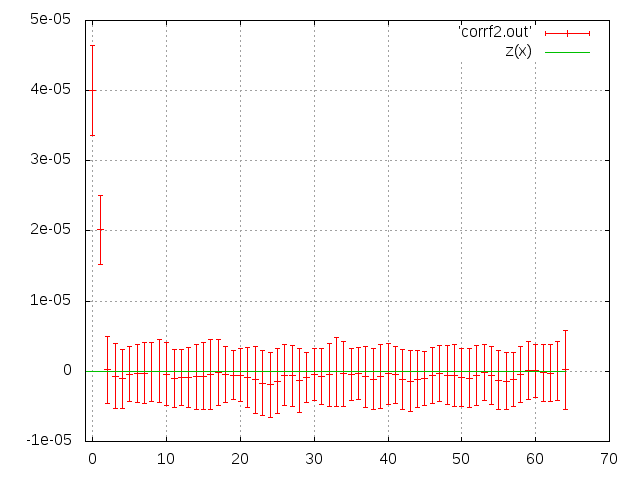}}
\subfigure{\includegraphics[scale=0.5]{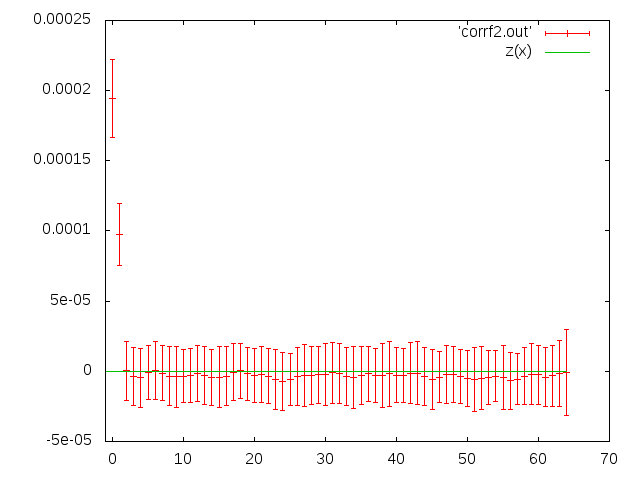}}
\caption[]{The connected 2--point function, $\langle\bm{\eta}_0\bm{\eta}_{|n-n'|}\rangle-\langle\bm{\eta}\rangle^2$, for the cubic superpotential, when the classical scalar potential has a unique minimum;   $g=0.9$ and $g=2.0$ and $m_\mathrm{latt}^2=0.0001$. In both cases they are $\delta-$functions, up to lattice artifacts (the point at $|n-n'|=1$) and the value at the origin is, indeed, equal to $g^2|m_\mathrm{latt}^2|/2$.} 
\label{2ptcubicw}
\end{figure}

For the cubic superpotential the corresponding results are displayed in figs.~\ref{1ptcubicw},~\ref{BiCcubicw} and~\ref{2ptcubicw}, for the case when the scalar potential possesses a unique minimum and the scalar is shifted so that this minimum is at the origin; and $g=0.9$, i.e. ``weak coupling''. 
The corresponding results for $g=2.0$, i.e. ``strong coupling'' are displayed in figs.~\ref{1ptcubics},
\begin{figure}[thp]
\includegraphics[scale=0.7]{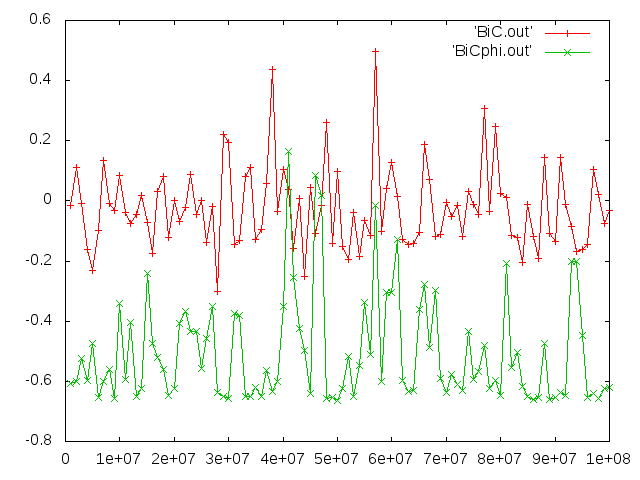}
\caption[]{The time series for the Binder cumulants, $-B_1(\bm{\eta})$ and $-B_1(\bm{\varphi})$, for the cubic superpotential, when $g=2.0$  and the classical scalar potential has a unique minimum. $m_\mathrm{latt}=0.0001$.} 
\label{BiCcubics}
\end{figure}
As expected, the 1--point function, $\langle\bm{\eta}\rangle\neq 0$ for $g=0.9$ and $g=5.0$, so supersymmetry is spontaneously broken. 

However, what's interesting is how ``weakly'' it appears to be broken: $\langle\bm{\eta}\rangle$ is very small in the scheme of things, whether at ``weak'' or ``strong'' coupling, as measured by $g$.

Here, too, supersymmetry is, nonetheless, relevant, because the noise field, still, describes a Gaussian field--just its 1--point function no longer vanishes. Its 2--point function is, still, ultra--local and its higher--point functions, that satisfy Wick's theorem, lead to relations between the correlation functions of the scalar, that are not at all obvious from the point of view of the classical action alone--they probe the relevance of the fluctuations. 

The validity of the identities, also, highlights that it isn't necessary to work with anything more than the Euclidian action of the commuting variables--nevertheless, the noise fields reveal the relevance of the anticommuting variables.

  \section{Conclusions}\label{conclusions}
In this paper we have provided concrete numerical evidence  that the fluctuations of a bath of additive white noise, with which a non--relativistic particle is in equilibrium, can be consistently described by its interaction with another particle, that moves in two, anticommuting dimensions, for every dimension of the target space that the ``ordinary'' particle probes; and that these dimensions are related through  worldline supersymmetry. 

What this amounts to is noticing that the existence of the canonical partition function, $Z_\mathrm{QM}$, implies that the Euclidian action is bounded from below and confines at infinity, therefore can be written as an integral of a perfect square
\begin{equation}
\label{Seclidsquare}
S_\mathrm{E}[\phi]=\int\,d\tau\,\frac{1}{2}\eta(\phi(\tau))^2
\end{equation}
Worldline supersymmetry means that the the fluctuations, described by the canonical partition function, 
\begin{equation}
Z_\mathrm{QM}=\int\,[\mathscr{D}[\phi]\,e^{-S_\mathrm{E}[\phi]}
\end{equation}
can be packaged as the contribution $|\mathrm{det}\delta\phi(\tau)/\delta\eta(\tau')|$, which means that $Z_\mathrm{QM}=1$. 

This holds whatever the nature of the fluctuations. For quantum mechanics, in particular,  worldline supersymmetry is not a property that must be added to quantum mechanics, to obtain ``supersymmetric quantum mechanics'', but is an intrinsic property thereof--far from being surprised at its presence, we should be surprised it hasn't been noticed till now. ``Supersymmetric quantum mechanics=quantum mechanics'', i.e. a tautology, since the quantum fluctuations are described by $(1,2,1)$ supermultiplets (i.e. 1 commuting, 2 anticommuting and 1 auxiliary fields). 

This is the most elementary supermultiplet; supermultiplets with different structure, i.e. $(n_B, n_F, n_A)\neq \mathrm{const}\times (1,2,1)$, but where, nonetheless, $n_F = n_B + n_A$, that implies closure, describe a  geometry of the target space that is, correspondingly, more intricate~\cite{Smilga:2013qy,Ivanov:2019gxo,Fedoruk:2015lza} and references therein. What's relevant is the closure relation itself, since the individual components themselves don't have intrinsic meaning.  

This raises the question of what happens, when supersymmetry is spontaneously broken, as is the case for the cubic superpotential. We propose that what this means is that additional degrees of freedom are necessary to complete the description, beyond the superpartners. The statement that the 1--point function of the noise field is non--zero means that the system isn't, in fact, closed.

That it is the coefficient of the linear term of the superpotential that controls whether supersymmetry is broken or not shouldn't be a surprise, of course; it illustrates that it is the boundary contribution(s) that ensure consistent closure and that this can't be achieved for all values of this coefficient (whereas this is the case for the quartic superpotential). 

It should be noted that there's been considerable recent activity in emergent supersymmetry in non--equilibrium systems~\cite{Gao:2017bqf,Mallick:2010su,Aron:2017spi,Gozzi:1983rk}, for instance. The difference from the approach described here is that, in the present paper, we have shown that supersymmetry is an intrinsic property of equilibrium systems.  

Away from equilibrium time translation invariance is broken. This, however, actually, can be understood as meaning that the global invariance isn't present--which can be, in turn, understood as signifying that the physical system in question, in fact, evolves in a non--trivial gravitational background! Which means that holography can be expected to be a useful tool for understanding the properties of such systems 
(cf.~\cite{Frassek:2019vjt,Facoetti:2019rab} for recent work in this direction, that does mention the stochastic approach, but doesn't push it to its logical conclusion, as is done here). 

Returning to equilibrium systems, we  would like, also, to stress,   that this (global) worldline supersymmetry doesn't have anything to do with any BRST symmetry, that has, often been mentioned in this context. The anticommuting fields, that resolve the fluctuations of the bath, aren't ghosts (that have been used in ref.~\cite{Deotto:2001sy}, for instance, for describing constraints).

 The way to render worldvolume supersymmetry manifest is by studying the identities of the noise field, expressed as a function of the scalar. The backreaction of the dynamics of the scalar on the fluctuations can be consistently described and the properties of the effective potential that does take the fluctuations into account can be described by the study of a local theory of a commuting field, that can be identified with the trajectory of a particle.The reason the position of a particle isn't a good observable, when fluctuations are  relevant, is that it becomes a component of a superfield and supersymmetric invariants are the only observables.

 We have studied some properties of the anharmonic oscillator, described by polynomial potentials, at the ``(semi)classical'' approximation, and have presented a way to probe the relevance of supersymmetry for such systems. Experiments with trapped ions or magnetic systems~\cite{Wineland:1997mg,Liu2011,Pientka} can, therefore, be used to study worldline supersymmetry and its breaking--which defines just what an ``open system'' actually means--quantitatively. They entail the control of non--linear susceptibilities for checking the identities these are expected to satisfy; the consistent coupling is that between a chiral multiplet and a vector multiplet, whose consequences in this context remain to be worked out (cf. ref.~\cite{DHoker:1983zea}).  They have been studied in the context of (super)conformal mechanics (cf.~\cite{Plyushchay:2018otc} for a recent review), in particular for applications to black holes~\cite{Galajinsky:2011xp,Orekhov:2016bpc}.

The present calculations show that the description of the dynamics of a particle, that explores a flat target space,  in a fluctuating environment,  not only realizes the slogan  ``all superparticles are BPS''~\cite{Mezincescu:2014zba,Mezincescu:2015apa,Baulieu:1993zm}--in the sense that the {\em classical} equations of motion are first order, since the Euclidian action can be written as the sum of squares--but that all such quantum particles are superparticles--even though supersymmetry may appear to be broken. 

Nonetheless there are subtleties, that slogans, of course, may seem to  miss--but that can, now, be elucidated:  A particle, that explores a flat target space, that has more than one dimensions, has a classical trajectory, labeled by $\phi^I(\tau)$, $I=1,2,\ldots,D$. The Langevin equation that describes its equilibrium dynamics is given by the expression(s):
\begin{equation}
\label{multiDlangevin}
\eta^I(\tau)=\frac{d\phi^I(\tau)}{d\tau}+\frac{\partial W}{\partial\phi^I(\tau)}
\end{equation}
That the target space is flat is expressed, beyond the fact that the coefficient of $\partial W/\partial\phi^I$ is a constant,  by the fact that the 2--point function of the noise field is given by the expression
\begin{equation}
\label{noise2ptD}
\left\langle\eta^I(\tau)\eta^J(\tau')\right\rangle=\nu\delta^{IJ}\delta(\tau-\tau')
\end{equation}
These properties imply that the classical, Euclidean,  action is given by the expression
\begin{equation}
\label{SclassD}
S[\phi^I]=\int\,d\tau\,\delta^{IJ}\left\{
\frac{1}{2}\dot{\phi}^I\dot{\phi}^J + \frac{1}{2}\frac{\partial W}{\partial\phi^I}\frac{\partial W}{\partial\phi^J}
\right\}
\end{equation}
since the mixed term
\begin{equation}
\label{mixedterm}
\delta^{IJ}\dot{\phi}^I\frac{\partial W}{\partial\phi^J}
\end{equation}
is a total derivative. 

This highlights that the scalar potential, $V(\phi^I)$ has a quite particular form:
\begin{equation}
\label{scalarpotD}
V(\phi^I)=\frac{1}{2}\delta^{IJ}\frac{\partial W}{\partial\phi^I}\frac{\partial W}{\partial\phi^J}
\end{equation}
that ensures, indeed, integrability of the motion, since it implies the BPS property, that leads to the reduction to first order. 

For one--dimensional target spaces the converse, also, holds: As mentioned in the introduction, any function $V(\phi)$, bounded from below, may be written as $(1/2)(W'(\phi))^2$--though $W'(\phi)$ need not be polynomial.  
Hence the slogan doesn't miss any subtleties, beyond that,  for multiple classical minima, tunneling leads to a quantum potential, whose form, defining a unique minimum,  is completely different from that of the (semi--)classical, polynomial expression (which is a lot, of course!). 

For higher dimensional target spaces, there are, however, additional,  subtleties: Given a function $W(\phi^1,\phi^2,\ldots,\phi^D)$, the scalar potential 
$V(\phi^1,\phi^2,\ldots,\phi^D)$ doesn't, necessarily,  describe integrable motion, even though the Euclidian action factorizes. 

The  reason the classical motion isn't, necessarily, integrable is that the classical vacua, for $D>1$, that are solutions of the system 
\begin{equation}
\label{classicalvacua}
\frac{d\phi^I}{d\tau}=-\frac{\partial W}{\partial\phi^I}
\end{equation}
can, for $D>2$, be chaotic attractors. Indeed this is one way to understand the scope of the Fayet--O'Raifeartaigh mechanism~\cite{Fayet:1975ki,ORaifeartaigh:1975nky} for supersymmetry breaking, that goes much further than the absence of real roots for the RHS of eq.~(\ref{classicalvacua}). 

One way to avoid them would be to consider equations of the form
\begin{equation}
\label{classicalvacua1}
\frac{d\phi^I}{d\tau}=-e^{IJ}(\phi)\frac{\partial W}{\partial\phi^J}
\end{equation}
 with the functions, $e^{IJ}(\phi)$ chosen so that the vector field on the RHS be divergence--free. This, still, wouldn't rule out Hamiltonian chaos, of course. 
 
Furthermore, obtaining the superpotential, $W$, from the scalar potential, $V$ involves solving the non--linear differential equation
\begin{equation}
\label{WfromV}
V(\phi^1,\phi^2,\,\ldots,\phi^D)=\frac{1}{2}\sum_{I=1}^D\left(\frac{\partial W}{\partial\phi^I}\right)^2
\end{equation}
This is the so--called ``eikonal equation''--however finding its solution, for $D>1$ is not at all straightforward  and just what the appropriate boundary conditions are is by no means obvious. 

This is, but one, reason, why the study of the correlation functions of the ``noise field(s)'', 
\begin{equation}
\label{noisefieldsDgt1}
\eta^I = \frac{d\phi^I}{d\tau}+e^{IJ}(\phi)\frac{\partial W}{\partial\phi^J}
\end{equation}
for general mechanical systems hasn't, to date, received the attention it deserves. Such considerations may provide new insights into ``swampland'' vs. ``landscape''~\cite{Ooguri:2016pdq}, since they can provide insights into the constrants that are involved in realizing reparametrization invariance of the worldline, i.e. worldline supergravity~\cite{vanHolten:2019fzy}. 
We hope to report on this subject in future work. Further applications to higher dimensional target spaces (that have their origins in the studies of the superparticle~\cite{Brink:1981nb,Brink:1981rt,Siegel:1983hh,Bergshoeff:1991hb})
are reviewed in~\cite{Ivanov:2019gxo,Fedoruk:2015lza} and it is of interest to elucidate the relation between the anticommuting variables described there and the fluctuations, as described by the Langevin equation--that will, necessarily, involve multiplicative noise. 

To summarize, we have shown that supersymmetry, more generally, is an ``emergent'' property of particle models--and, also, more generally, in that it is the property that ensures the ``coordinate independence'' in the space of fields, when choosing which fields resolve the bath and which resolve the dynamics one is interested in (cf. also~\cite{Gies:2017tod,Li:2017dkj,Zhao:2017bhw}). 
 In the process, we have provided a new way of classifying supersymmetric models, going beyond~\cite{Smilga:2013qy}.

What hasn't been appreciated to date is that equilibrium of a physical system--that is described in terms of particles--with a bath, means that there exist other particles, whose 
collective dynamics is represented by the properties of the bath. What Parisi and Sourlas stressed is that the bath can be described by anticommuting variables, if the dynamics of the canonical partition function is described by commuting variables--and that the equilibrium can be described by invariance under supersymmetric transformations. In this way can the system be consistently closed~\cite{Hartle:1992bf}.

However, it turns out that supersymmetry appears in another, totally different way, in the dynamics of a single particle, namely in the dynamics of the relativistic, spinning, particle~\cite{Brink:1976uf}. The anticommuting partners, $\psi^\mu(\tau)$, of the position $\phi^\mu(\tau)$, in this case, are identified with the spin degrees of freedom, that describe classical dynamics, which  requires a phase space with both kinds of coordinates--this has been extensively studied in ref.~\cite{Casalbuoni:1975bj,Berezin:1975md,Berezin:1976eg,Bachas:1978fs}. (In fact the relativistic spinning particle realizes worldline supergravity, as noted in ref.~\cite{Brink:1976uf}.) 
The generalization to field theories, i.e. of systems of an indefinite number of particles, was worked out in~\cite{Ferrara:1974pu}. What's interesting is that  ordinary gauge theories, coupled to Majorana spinors, that transform in the adjoint representation of the gauge group, in fact, are supersymmetric. The question that wasn't addressed was what could be the physical realization of such particles. There was some discussion, e.g.~\cite{Barducci:1980xk}, that hasn't been pursued, however, to a definite conclusion. We hope that the clarification provided in the present study will lead to experiments in modern condensed matter systems, that will provide illustrations of the relevance of supersymmetry in ``usual'' quantum systems and in fluctuating systems, more generally. 

What is important to keep in mind is that the separation of the anticommuting degrees of freedom into those that can be assigned to the spin variables and those that can be assigned to the fluctuations of the commuting degrees of freedom (and vice versa!) for instance, depends on how the corresponding constraints are imposed~\cite{Brink:1976uf}. The appearance of anomalies has only recently been studied~\cite{Papadimitriou:2017kzw,Katsianis:2019hhg} and it will be interesting to study how their avatars appear in the simpler context of particle mechanics, that can be related to traps and magnetic moments.

The implication of the present article is that the quantum dynamics of a spinning particle is, fully, described by {\em three} supersymmetries: The supersymmetry that relates the position and the spin of the particle; the supersymmetry that relates the position and the anticommuting variables, that describes its fluctuations; and the supersymmetry that relates the spin degrees of freedom and {\em its} fluctuations. Recent progress in describing the fluctuations in magnetic materials should allow measurements of the non--linear susceptibilities, whose relations are the stochastic identities of supersymmetry.  (Attempts to compute the equal--time susceptibilities are discussed in ref.~\cite{Tranchida:2016xkj,Nussle:2017gwl,Nussle:2019kwe}.)

Let us close with how these results can be extended to relativistic theories and to gauge theories. 

To describe relativistic theories means describing target space fermions. Such theories were, also, discussed in ref.~\cite{parisi_sourlas}. The generalization of the Langevin equation, applicable to Wess--Zumino models,  is given by the expression 
\begin{equation}
\label{targetspace}
\eta^I = \sigma_{IJ}^\mu\frac{\partial\phi^J}{\partial x^\mu}+\frac{\partial W}{\partial\phi^I}
\end{equation}
where the $\sigma^\mu$ generate a Clifford algebra, $\{\sigma^\mu,\sigma^\nu\}=2\delta^{\mu\nu}$ in Euclidian signature. The case of a two--dimensional, flat, target space was studied, using the approach described here in ref.~\cite{Nicolis:2017lqk}, where it was found that, indeed, it's not necessary to take into account the stochastic determinant-or its sign--explicitly, but to check that the stochastic identities are satisfied, using the canonical partition function.  These results highlight that (a) the canonical partition function does, indeed, describe all the fluctuations and (b) that it isn't necessary to deal withe fermions on the lattice directly, in this case, either. The properties of the fermions are to be deduced from those of the scalars, using the supersymmetry transformations. That's what ``consistent closure'' of the system entails, in practice.

Gauge theories have the complication that the target space they define--the group manifold--isn't globally flat, so the Langevin equation for motion on the group manifold is more complicated~\cite{Gausterer:1987uz}, namely describes multiplicative noise. 
 (The stochastic quantization procedure, that reproduces the perturbative expansion about free fields~\cite{Floratos:1982xj,Floratos:1983nk}, works with fields that take values in the algebra, i.e. the tangent space to the group manifold.) How to implement it efficiently for such cases remains to be worked out. Toy models for such an approach are, indeed, magnetic moments, embedded in an elastic medium and subject to multiplicative noise~\cite{Tranchida:2016xkj,Nussle:2017gwl,Nussle:2019kwe}.

{\bf Acknowledgements:} It's a pleasure to thank C. P. Bachas,  E. G. Floratos, J. Iliopoulos,  E. Ivanov, A. Smilga and T. N. Tomaras for their constructive criticism,   M. Axenides, X. Bekaert, B. Boisseau and H. Giacomini  for helpful suggestions and C. Winterowd and S. Zafeiropoulos for their interest in this project and for teaching the Hybrid Monte Carlo techniques that will be instrumental in taking this project to the next level.   The warm hospitality of the LP(T)ENS and the NRCPS ``Demokritos'' is gratefully acknowledged.

\bibliographystyle{utphys}
\bibliography{SUSY}

\end{document}